\documentclass{elsart}

\journal{Journal of Computational Physics}

\usepackage{latexsym,amssymb,graphics,graphicx,deluxetable}

\def\eps@scaling{1.0}% 
\newcommand\epsscale[1]{\gdef\eps@scaling{#1}}% 
\newcommand\plotone[1]{% 
 \centering 
 \leavevmode 
 \includegraphics[width={\eps@scaling\textwidth}]{#1}% 
}% 
\newcommand\plottwo[2]{% 
 \centering 
 \leavevmode 
 \textwidth=.45\textwidth 
 \includegraphics[width={\eps@scaling\textwidth}]{#1}% 
 \hfil
 \includegraphics[width={\eps@scaling\textwidth}]{#2}% 
}% 
\newcommand\plotfiddle[7]{% 
 \centering 
 \leavevmode 
 \vbox\@to#2{\rule{\z@}{#2}}% 
 \includegraphics[% 
  scale=#4, 
  angle=#3, 
  origin=c 
 ]{#1}% 
}% 
\usepackage[square,comma,numbers,sort]{natbib}
\renewcommand\bibsection{\section*{References}}
\newif\ifAMStwofonts
\ifAMStwofonts \else % If no AMS fonts

\fi

\newcommand{\pd}[2]{ { \partial {#1} \over \partial {#2} } }
\newcommand{\pdd}[2]{ { \partial^2 {#1} \over \partial {#2}^2 } }

\newcommand{\vect}[1]{\mathbf{#1}}   % Vector as bold
\newcommand{\sfrac}[2]{\,{}^{#1}\!/_{#2}}

\newcommand{\beq}{\begin{equation}}
\newcommand{\eeq}{\end{equation}}
\newcommand{\bdm}{\begin{displaymath}}
\newcommand{\edm}{\end{displaymath}}

\begin{document}

\begin{frontmatter}

% Title, authors and addresses

% use the thanksref command within \title, \author or \address for footnotes;
% use the corauthref command within \author for corresponding author footnotes;
% use the ead command for the email address,
% and the form \ead[url] for the home page:
% \title{Title\thanksref{label1}}
% \thanks[label1]{}
% \author{Name\corauthref{cor1}\thanksref{label2}}
% \ead{email address}
% \ead[url]{home page}
% \thanks[label2]{}
% \corauth[cor1]{}
% \address{Address\thanksref{label3}}
% \thanks[label3]{}

\title{Computation of Fluid Flows in Non-inertial Contracting, Expanding, and 
Rotating Reference Frames}

% use optional labels to link authors explicitly to addresses:
% \author[label1,label2]{}
% \address[label1]{}
% \address[label2]{}

\author[Address]{Alexei Y. Poludnenko\corauthref{cor}},
\corauth[cor]{Corresponding author.}
\ead{apol@uchicago.edu}
\author[Address]{Alexei M. Khokhlov}
\ead{ajk@oddjob.uchicago.edu}

\address[Address]{Center for Astrophysical Thermonuclear Flashes,
Department of Astronomy and Astrophysics, University of Chicago,
5640 South Ellis Avenue, Chicago, IL 60637}

\begin{abstract}
We present the method for computation of fluid flows that are
characterized by the large degree of expansion/contraction and in
which the fluid velocity is dominated by the bulk component associated
with the expansion/contraction and/or rotation of the flow. We
consider the formulation of Euler equations of fluid dynamics in a
homologously expanding/contracting and/or rotating reference
frame. The frame motion is adjusted to minimize local fluid
velocities. Such approach allows to accommodate very efficiently large
degrees of change in the flow extent. Moreover, by excluding the
contribution of the bulk flow to the total energy the method
eliminates the high Mach number problem in the flows of interest.  An
important practical advantage of the method is that it can be easily
implemented with virtually any implicit or explicit Eulerian
hydrodynamic scheme and adaptive mesh refinement (AMR) strategy.

We also consider in detail equation invariance and existence of
conservative formulation of equations for special classes of
expanding/contracting reference frames. Special emphasis is placed on
extensive numerical testing of the method for a variety of reference
frame motions, which are representative of the realistic applications
of the method. We study accuracy, conservativity, and convergence
properties of the method both in problems which are not its optimal
applications as well as in systems in which the use of this method is
maximally beneficial. Such detailed investigation of the numerical
solution behavior is used to define the requirements that need to be
considered in devising problem-specific fluid motion feedback
mechanisms.

\end{abstract}

\begin{keyword}
Expanding flows; Contracting flows; Rotating flows; Moving frame; Moving
mesh; Hydrodynamics
\PACS
\end{keyword}
\end{frontmatter}

% main text

\section{Introduction}
%-------------------------------------------------------------------------------
\label{Intro}

Many fluid dynamical problems, particularly the ones of high
importance in astrophysics and cosmology, have the following two key
characteristics: (1) the fluid undergoes a very large degree of
expansion/contraction on its evolutionary timescale; (2) the flow
consists of the global component, associated with the overall
expansion/contraction and, if present, rotation of the system and the
superimposed local fluid velocity field, with the global flow velocity
being much larger than the local fluid velocity or the local sound
speed. Additionally, such systems are often characterized by the
linear distribution of total velocity along the radius, sometimes
referred to as the Hubble flow by analogy with cosmology.
Astrophysical examples include, but are not limited to, stellar core
collapse, supernova explosions (SNe), star and galaxy formation, as
well as cosmological models. One of the most prominent examples of
such systems in terrestrial applications is the inertial confinement
fusion (ICF) problem. Compression or expansion of matter in those
problems may reach many orders of magnitude.

Problems with large degree of expansion/contraction are
computationally difficult. Local features of the flow in those
problems are significantly compressed, expanded, and advected over
large distances. This puts extreme demands on numerical resolution and
on the quality of numerical advection algorithms. For the rotating
fluid large compression or expansion may also lead to large numerical
errors in conservation of angular momentum.

Three different approaches in the context of Eulerian formalism can be
used to overcome some of those computational difficulties: (1)
adaptive mesh refinement (AMR), (2) computations on a moving mesh
(MM), and (3) computations in a moving (non-inertial) reference
frame (MRF). The first two approaches are being widely used to address
the first characteristic of the flows, mentioned above, namely to
accommodate large degrees of flow expansion or contraction. In an AMR
approach (e.g., see \citep{BergerColella,Khokhlov}), the computational
mesh is refined or derefined to counteract respectively flow
contraction or expansion thus maintaining numerical resolution of the
features of interest. One representative example is the AMR
calculation of star formation in the early Universe (see
\citep{AbelNorman} and references therein). As opposed to increasing
or decreasing the number of cells following the flow in the AMR
approach, in the MM techniques mesh lines are moved continuously to
minimize the relative motion of the fluid with respect to the
computational grid \citep{HartenHyman,Winkler,FazioLeVeque} thereby
providing expansion or contraction of the grid in accordance with the
motion of the flow. This general method also includes a large class of
techniques referred to as arbitrary Lagrangian-Eulerian method (ALE)
\citep{Hirt,Anderson} in which the mesh is deformed to follow the
fluid flow. A recent astrophysical example is the use of a traditional
moving mesh technique \citep{Winkler} to follow explosion of a type Ia
SNe to the stage of free ballistic expansion \citep{Roepke,
RoepkeHillebrandt}.

The AMR and MM approaches are fundamentally similar in that they both
work with fluid quantities defined in a stationary inertial reference
frame. The only difference is that in the AMR approach the fluid moves
through a stationary mesh and an additional interpolation is required
only when the mesh is refined or derefined. In the MM approach
coordinate transformation between the physical space and the
computational domain is provided and at every time step the formal
correspondence (transformation map) between computational and physical
coordinates is established. Thus fluid quantities must be
re-interpolated onto a new mesh every time step either explicitly via
an Eulerian step plus re-map, or implicitly by modifying fluxes
through boundaries of computational cells. Since fluid variables are
defined in a stationary frame they are not affected by mesh movements.
While the AMR and MM approaches are capable of addressing the first
characteristic of the flows discussed above, this key property of
those methods makes them inefficient in modeling flows dominated by
the global component associated with expansion/contraction and/or
rotation. In such systems practically all of the kinetic energy is due
to the global flow, thus the ratio of thermal to kinetic energy can be
extremely small. Consequently, solving for the total energy in the
inertial reference frame results in large errors in thermal energy
and, thus, pressure. This situation is well known in hydrodynamics as
a high Mach number problem and various methods have been employed to
solve it in different contexts (e.g., see
\citep{TracPen} and references therein).

The goal of the MRF approach is to address that problem while
preserving the efficiency of the MM method in accommodating large
degrees of the flow expansion/contraction. In the MRF method the full
reference frame transformation is performed, as opposed to only the
coordinate transformation in the MM approaches. Fluid velocities and
total energy are defined with respect to a comoving reference frame in
which thermal and local kinetic energies are comparable in magnitude
thereby eliminating the high Mach number problem. Computational mesh
in this approach has two distinct functions. It defines the boundaries
of computational cells and, at the same time, represents a reference
frame. Consequently, the equations of fluid dynamics must be modified
to include the effects of frame expansion, acceleration, as well as
forces arising due to the reference frame rotation.

The area of widest use of the MRF approach in astrophysics, albeit in
a somewhat simplified sense, are cosmological simulations. In those
simulations the terms accounting for expansion of the Universe are
known a priori and are explicitly added as source terms to the fluid
equations \citep{AbelNorman}. In \citep{TracPen} a collapse of a
cosmological pancake was considered in a non-inertial reference frame
the motion of which was implicitly defined by self-gravity of the
pancake structure itself. It is impossible to pick a single "best"
numerical approach to solving all fluid dynamics problems. The right
choice must depend on the problem in question and often it is a
compromise between accuracy, flexibility, ease of applicability, and
code availability. The approaches discussed above can be and are often
used in combination. For example, simulations of galaxy formation
routinely combine a MRF approach which accounts for global expansion
of the Universe with the AMR or MM approaches used for a more accurate
treatment of structure formation on smaller scales.

In this paper we investigate the applicability of the MRF approach and
its combination with AMR to astrophysical problems involving
expanding / contracting and rotating objects such as collapsing stellar
cores and supernovae. Such systems exhibit the two key characteristics
discussed above. We also have in mind applications to implosion of ICF
targets ablated by laser radiation.

We consider non-inertial reference frames which expand or contract
spherically-symmetrically with respect to an inertial laboratory
frame. A solid (non-differential) rotation of the frame is also
allowed. In many practical cases this may be enough to compensate for
the bulk motion associated with the explosion of a star or implosion
of an ICF target. We work on a premise that peculiar motions, local
deformations, and sharp features, e.g., shocks, contact and material
discontinuities, and reaction fronts, present in the flow can be
better treated using the AMR method applied in a moving frame. Our
goal in this work is to investigate general numerical properties of
the method which involves computations in such a non-inertial
reference frame. In particular, our emphasis is on the accuracy,
conservativity, and convergence properties of the method under
different types of reference frame motions. Such properties can then
be used in devising problem-specific feedback mechanisms that define
specific reference frame motions based on the fluid flow evolution.

In \S~\ref{EulerEquations} we discuss the general transformation to
the moving non-inertial reference frame of the type discussed above,
we write down the transformed set of fluid equations and we also
discuss the invariance of the original equations and the existence of
conservative formulation. In \S~\ref{NumMethod} we discuss the
numerical method used in solving the transformed set of fluid
equations. In \S~\ref{NumTests} we present results of numerical
tests. We use two completely independent AMR codes - ALLA and
AstroBEAR \citep{Khokhlov,Poludnenko}, which differ in their
hydrodynamic integration schemes and in their AMR approach (cell-based
vs. grid- or patch-based). The testing is twofold. On one hand, we
present a series of tests which are not the optimal applications of
this method and which are designed to stress the method. They include
the strong point explosion (Sedov blast wave) and the converging shock
(Guderley blast wave). However, those tests give a very good idea of
the limitations of the MRF method which must be always kept in mind
while using the MRF approach in real applications. We also consider a
series of tests for which this method is extremely well suited. Those
tests are based on the expansion of a non-rotating and a rotating
sphere into vacuum and isentropic expansion of a uniform pressure
field with embedded density structure. Discussion and conclusions are
given in \S~\ref{Discussion}, where we also discuss performance
comparison of the method presented here and the moving mesh approach
using as an example the moving mesh implementation of the Zeus-MP code
\citep{Zeus}. Finally, Appendix A presents the special case of
transformed fluid equations in cylindrical coordinates in the absence
of rotation.

\section{Equations of Fluid Dynamics in an Expanding/Contracting and
Rotating Reference Frame}
%-------------------------------------------------------------------------------
\label{EulerEquations}

\subsection{General formalism}
\label{GeneralFormalism}

We start with an inertial Cartesian (laboratory) reference frame
$\mathbf{X} = \{\mathbf{r},t \}$. Euler equations of fluid dynamics in
this frame are
\begin{eqnarray}
\pd{\rho}{t} + \nabla \cdot (\rho \vect{u} )    & \quad = \quad & 0,
\label{Euler1} \\
\pd{\rho \vect{u}}{t} + \nabla \cdot ( \rho \vect{u} \otimes \vect{u}) 
				+ \nabla P      & \quad = \quad & 0,
\label{Euler2} \\
\pd{E}{t} + \nabla \cdot \big((E+P) \vect{u}\big) & \quad = \quad & 0,
\label{Euler3}
\end{eqnarray}
where $\rho$ is mass density, $\vect{u}$ - fluid velocity, $P$ -
pressure, $E = \rho \, e + \frac{1}{2}\rho u^2$ - energy density, and
$e$ - internal energy per unit mass.

Next we consider a non-inertial reference frame $\mathbf{\tilde X} =
\{\vect{\tilde r},\tau \}$, which rotates and homologously expands or
contracts with respect to $\mathbf{X}$. The frame $\mathbf{\tilde X}$
is defined by the transformation $\mathbf{\Lambda}: \mathbf{X}
\mapsto \mathbf{\tilde X}$
\beq
\mathbf{\Lambda} = \left\{ \begin{array}{lcl}
\vect{\tilde r}  & = &  \displaystyle a^{-1} \left( \vect{r} - 
\int_0^t \vect{\omega}(t) dt \times \vect{r} \right), \\
\tau        & = & \displaystyle \int_0^t \frac{dt}{a^{\beta+1}},
\end{array} \right.
\label{Transform}
\eeq
where $\beta$ is a constant. We require the transformation
$\mathbf{\Lambda}$ to be non-singular, therefore we assume that the
\emph{scale factor} $a(t)$ is a smooth non-vanishing
twice-differentiable function of time only. Angular velocity of the
frame $\vect\omega(t)$ is assumed to be a smooth once differentiable
function of time. Here we consider only the cases of solid body
rotation, i.e., the cases when $\vect{\omega}$ is only a function of
time and not spatial coordinates, and we assume that $\vect{\omega}$
does not change its spatial orientation. Note, that $d\tau =
dt/a^{\beta + 1}$. The inverse transformation is
\beq
\mathbf{\Lambda}^{-1} = \left\{ \begin{array}{lcl}
\vect{r}  & = &  \displaystyle a(\tau) \left( \tilde\vect{r} - 
\int_0^\tau \vect{\Omega}(\tau) d\tau \times \tilde\vect{r} \right), \\
t         & = & \displaystyle \int_0^\tau a^{\beta + 1}(\tau) d\tau,
\end{array} \right.
\label{InverseTransform}
\eeq
where we introduced the \emph{effective angular velocity}
\beq
\vect{\Omega} = a^{\beta+1} \vect{\omega} = a^{\beta+1} \frac{d\vect{\phi}}{dt} =
\frac{d\vect{\phi}}{d\tau},
\label{Omega}
\eeq
which describes the angle swept by the reference frame $\mathbf{\tilde
X }$ per unit computational time. 

Hereafter, quantities with the \emph{tilde} sign refer to the
reference frame $\mathbf{\tilde X}$. For simplicity we will also be
referring to the stationary laboratory frame $\mathbf{X}$ as the
\emph{physical frame} and to time $t$ as the \emph{physical
time}. Since typically in the numerical tests discussed below the
computational grid is associated with the frame $\mathbf{\tilde X}$ we
will be referring to this frame as the \emph{computational frame} and
to time $\tau$ as the \emph{computational time}.

The transformation $\mathbf{\Lambda}$ implies the decomposition of the
velocity field $\vect{u}$, present in the frame $\mathbf{X}$, into the
sum of the bulk velocity associated with the global
expansion/contraction and rotation, and the superimposed local
velocity field $\vect{\tilde u}$ in the frame $\mathbf{\tilde X}$
\beq
\vect{u} = \vect{\Omega} \times \vect{\tilde r}
	 + a^{-\beta}\frac{d\ln a}{d\tau} \vect{\tilde r}
	 + a^{-\beta}\vect{\tilde u}.
\label{newvel}
\eeq
Thus $\mathbf{\Lambda}$ provides a homologous spatial transformation
ensuring that if matter expands/contracts spherically and rotates with
velocity $\vect{u'} = \vect{\omega}\times \vect{r} + H(t)\vect{r}$ in
the stationary frame $\mathbf{X}$, then that matter will be at rest in
the computational reference frame $\mathbf{\tilde X}$. The scale
factor $a(t)$ can be defined by $\dot{a}(t)/a = H(t)$\footnote{In
cosmological models $H(t)$ is referred to as the \emph{Hubble
parameter}.}, where $\dot a$ is the derivative with respect to
physical time and not the new transformed time $\tau$. Converting to
$\tau$ we get
\beq
\dot a = \frac{da}{d\tau} \frac{d\tau}{dt} = \frac{1}{a^{\beta}}\frac{d\ln a}{d\tau}.
\label{dot_a}
\eeq
Similarly, we note another useful relation
\beq
\ddot{a} = \frac{d\dot{a}}{d\tau}\frac{d\tau}{dt}
         = \frac{1}{a^{2\beta+1}}\Bigg\{\frac{d^2\ln a}{d\tau ^2}
	 - \beta\left( \frac{d\ln a}{d\tau}\right)^2\Bigg\}.
\label{ddot_a}
\eeq
Via the transformation $\mathbf{\Lambda}$ we, in effect, introduced
the scaling of length and time. We have the third independent physical
quantity, namely mass, the scaling of which we introduce via the
density field
\beq
\tilde \rho(\tilde \vect{r},\tau) = a^{\alpha}\rho.
\label{rhotr}
\eeq
Here $\alpha$, as well as $\beta$ in (\ref{Transform}), are the
scaling parameters.

In the reference frame $\mathbf{\tilde X }$ the Euler equations
(\ref{Euler1}) - (\ref{Euler3}) become
\begin{eqnarray}
  \pd{\tilde\rho}{\tau} & \, + \, & \tilde\nabla \cdot (\tilde\rho \, \vect{\tilde u}) =
\left(\alpha  - \nu \right) \frac{d\ln a}{d\tau} \tilde\rho,
\label{EulerTr1} \\
  \displaystyle \pd{\tilde\rho \, \vect{\tilde u}}{\tau} & \, + \, &
  \tilde \nabla \cdot (\tilde\rho \, \vect{\tilde u} \otimes \vect{\tilde u})
+ \tilde \nabla \tilde P = (\alpha - \nu + \beta - 1) \frac{d\ln a}{d\tau}
  \tilde\rho \, \vect{\tilde u} - \nonumber \\
&  & \displaystyle \Bigg\{\frac{d^2\ln a}{d\tau ^2}
-\beta\left( \frac{d\ln a}{d\tau}\right)^2\Bigg\}
  \tilde\rho \, \vect{\tilde r} - \tilde\rho \Bigg[ \vect{\Omega} \times 
  ( \vect{\Omega} \times \vect{\tilde r} ) + \nonumber \\
&  & \displaystyle \left( \frac{ d\ln \Omega}{d\tau} +
  \left(1 - \beta\right) \frac{d\ln a}{d\tau}\right) \, 
  \vect{\Omega}\times\vect{\tilde r} \Bigg] - 2 \vect{\Omega} \times \tilde\rho \vect{\tilde u},
\label{EulerTr2} \\
  \displaystyle \pd{\tilde E}{\tau} & \, + \, &
  \tilde \nabla\cdot \bigg( \vect{\tilde u} \left( \tilde E + \tilde P \right) \bigg) =
  \frac{d\ln a}{d\tau} \bigg[\left(\alpha - \nu + 2\beta \right) \tilde E
  - \nu\tilde P -\tilde \rho \tilde u^2 \bigg] - \nonumber \\
&  & \displaystyle \Bigg\{\frac{d^2\ln a}{d\tau ^2}
-\beta\left( \frac{d\ln a}{d\tau}\right)^2\Bigg\}
  \left( \tilde\rho \vect{\tilde u} \cdot \vect{\tilde r} \right)
- \left( \tilde\rho \vect{\tilde u} \cdot \vect{\Omega} \right)
  \left( \vect{\tilde r} \cdot \vect{\Omega} \right)
+ \Omega^2 \left( \tilde\rho \vect{\tilde u} \cdot \vect{\tilde r} \right) - \nonumber \\
&  & \displaystyle \left( \frac{ d\ln\Omega}{d\tau}
+ \left( 1 - \beta \right)\frac{d\ln a}{d\tau} \right)
  \tilde\rho \vect{\tilde u} \cdot \left( \vect{\Omega}\times\vect{\tilde r} \right),
\label{EulerTr3}
\end{eqnarray}
where $\tilde E = \tilde\rho \tilde e + \frac{1}{2} \tilde\rho \tilde
u^2$, and $\tilde \nabla$ indicates differentiation with respect to
spatial coordinates $\vect{\tilde r}$. Hereafter, $\nu$ is the
dimensionality parameter of the problem. The transformed pressure and
internal energy fields have the form
\beq
\tilde P(\tilde \vect{r},\tau) = a^{\alpha + 2\beta}P; \quad
\tilde e(\tilde \vect{r},\tau) = a^{2\beta}e;
\label{Ptr}
\eeq
The first terms on the right-hand side of eqs. (\ref{EulerTr1}) -
(\ref{EulerTr3}) are associated with expansion/contraction per se of
the reference frame, while the second terms represent the effects of
the frame acceleration. The remainder of the terms in eqs.
(\ref{EulerTr2}) and (\ref{EulerTr3}) describe the effects of the
frame rotation. In particular, the first term in square brackets in
eq. (\ref{EulerTr2}) is the centrifugal force while the second term is
related to the unsteadiness of the frame rotation and, finally, the
last term on the right-hand side of the equation is the Coriolis force.

The only two parameter functions that have not been defined yet are
the scale factor $a(t)$ and the angular velocity $\vect{\omega}(t)$.
Typically, the choice of these functions is problem-specific and their
temporal evolution is governed by fluid motions in the system. Thus a
realistic application of the method presented here requires a feedback
mechanism with appropriate filtering that will translate complex
multidimensional fluid motions into smooth functions $a(t)$ and
$\vect{\omega}(t)$. Since in this work we are primarily concerned with
the general properties of the method we do not consider such
feedback. Instead, we limit the discussion only to such types of
computational frames the motion of which with respect to the inertial
frame is predefined, e.g, via an analytic prescription. We chose a set
of computational frames that covers a large range of possible motions
and that is rather numerically challenging. This allows us to
illustrate below accuracy, stability, and conservativity properties of
our approach which can be later used in devising the problem-specific
feedback filters.

The method presented here is naturally able to accommodate large
degrees of expansion or contraction of fluid flows.  However, as it
was discussed in \S~\ref{Intro}, the second key characteristic of the
flows, which we are interested in modeling, is the fact that the fluid
velocity associated with the global flow due to its
expansion/contraction and, if present, rotation greatly exceeds both
the local peculiar velocities and the sound speed. In the context of
eq. (\ref{newvel}) this statement can be written as
\beq
|\tilde\vect{u}| \ll a^{\beta}|\vect{\omega}\times \vect{r^*} + H(t)\vect{r^*}|;
\quad \tilde c \ll a^{\beta}|\vect{\omega}\times \vect{r^*} + H(t)\vect{r^*}|.
\label{FlowCondition}
\eeq
where $H(t) = \dot{a}(t)/a$ and $r^*$ defines the maximum extent of
the flow, where the largest bulk flow velocities exist. Note that
because of the relations (\ref{rhotr}) and (\ref{Ptr}) the local sound
speed $c$ is transformed as $\tilde c = a^{\beta} c$. As an example,
consider Fig.~\ref{SphereEnergyEvolution} representing the increase of
kinetic energy in the case of expansion of a non-rotating and a
rotating sphere into vacuum. We discuss these two tests in detail in
\S~\ref{TestTypes}. As the sphere material is accelerated to its
terminal expansion velocity, kinetic energy rapidly approaches total
energy and then for more than 75\% of the total simulation run time in
the non-rotating case and more than 90\% of the time in the rotating
case it constitutes over 99\% of the total energy. Practically all of
that kinetic energy is associated with the global fluid flow. Thus in
the inertial frame thermal component is only a small fraction of the
total energy and solving for the latter would result in large errors
in pressure and, consequently, the overall solution. On the other
hand, in the computational frame $\mathbf{\tilde X}$, which closely
follows global fluid motions, thermal and local kinetic energies are
comparable in magnitude thereby eliminating the high Mach number
problem and resulting in much smaller errors in pressure.

The method presented here is not equally efficient in treating all
fluid flows that undergo substantial expansion/contraction in the
course of their evolution and exhibit property (\ref{FlowCondition}).
Consider shock strength $\Pi = \Delta P/\rho_0 c^2_0$, where $\Delta
P$ is pressure jump over the shock, $\rho_0$ is pre-shock fluid
density and $c_0$ is pre-shock sound speed. It is an invariant of the
transformation $\mathbf{\Lambda}$ and the associated transformations
(\ref{rhotr}) and (\ref{Ptr}). In practice, consideration of such
invariant quantities can define the extent of applicability of the
method presented here and its efficiency in a given application. Next
consider a problem involving propagation of the global shock into
stationary medium, e.g., due to a strong point explosion. In a
co-expanding frame $\mathbf{\tilde X}$, the velocity of which was
adjusted to the shock speed, the shock will be stationary. However,
its strength $\Pi$ must be maintained and that is achieved by the
opposing velocity gradient $\vect{\tilde u} = -\dot{a}\vect{\tilde
r}$, which is the direct result of the application of the velocity
transformation (\ref{newvel}) to the zero velocity field of the
ambient material. In general, any regions of material, which is
stationary in the inertial reference frame and which is dynamically
important, will have such velocity gradients which may be quite
large. This can limit the computational time step. Therefore, such
problems do not constitute an optimal class of applications for this
method. On the other hand, expanding or collapsing environments, in
which ambient conditions are vacuous or dynamically unimportant so
that the ambient fluid can be set to be stationary in the
computational frame, or in which the computational domain contains the
interior of the expanding or collapsing flow represent the class of
applications most suited for this method. Finally, it is always
advantageous to consider rapidly rotating flows in the co-rotating
frame. Below we consider examples of both types of problems.

Finally, it should be noted that while the system (\ref{Euler1}) -
(\ref{Euler3}), and consequently the transformed system
(\ref{EulerTr1}) - (\ref{EulerTr3}), are the most general sets of
equations, there are special cases, for example, employing specific
symmetries of the problem at question, that may be of interest in
various applications. One such case, namely the case of cylindrical
symmetry, is considered in Appendix A. There may also be more general
forms of the scale factor $a$ and the angular velocity of the frame
$\vect{\omega}$ that may be useful. One important generalization is
the scale factor that depends not only on time but also on radial
distance $a(r,t)$. This would allow one to isolate certain regions of
the domain that would be comoving with the global flow but that would
not experience expansion and, therefore, would not suffer from the
loss of numerical resolution in physical space. However, such
transformations are outside the scope of this work.

\subsection{Equation invariance and conservative formulation}
\label{EqnInvariance}

The first important question that has to be answered regarding eqs.
(\ref{EulerTr1}) - (\ref{EulerTr3}) is whether there exists their
conservative formulation. Traditionally it is considered beneficial in
the numerical hydrodynamics to work with the conservative formulation
of Euler equations. In the moving mesh approach, which relies on the
coordinate remapping instead of a true reference frame transformation,
it is always possible for any coordinate transformation to cast Euler
equations in a ``conservative'' form
\beq
\left( J q \right)_t + \tilde \nabla \cdot \tilde F = 0,
\label{MovMeshLaw}
\eeq
with a new vector of conserved quantities $q^* = Jq$, where $J$ is the
Jacobian of the coordinate transformation, and some modified flux
functions $\tilde F_i$. Indeed, primitive variable fields, including
velocities which are always defined in the inertial frame
$\mathbf{X}$, are not changed by the coordinate transformation.
Therefore, the Hamiltonian structure of the system remains invariant.
Consequently all conserved quantities are preserved and only their
volumetric densities are affected due to the rescaling of length
introduced by the coordinate transformation, which is reflected in the
Jacobian factor. When the true reference frame, and not just
coordinate, transformation is employed, e.g., the transformation
$\mathbf{\Lambda}$, that immediately changes the structure of Action
and the Hamiltonian of the system, which now attain some generally
non-trivial explicit temporal dependence due to the functions $a(t)$
and $\vect{\omega(t)}$. In particular, the Hamiltonian in the frame
$\mathbf{\tilde X}$ takes the form (see also \citep{Salmon})
\begin{eqnarray}
\tilde H & \, = & \, \int\int\int d\vect{\tilde x}\left(\tilde \rho \tilde e + 
\frac{1}{2}\tilde \rho \tilde u^2\right) =
  a^{\alpha+2\beta-\nu}\int\int\int d\vect{x}\Bigg(\rho e
+ \frac{1}{2}\rho u^2 - \nonumber \\
& & \frac{1}{2}\rho\Bigg\{2a^{\beta} \left(\vect{\omega}\times \vect{r}\right)\cdot
  \bigg(\vect{u} - \frac{a^{\beta}}{2}(\vect{\omega}\times \vect{r})\bigg)
+ 2\frac{\dot{a}}{a}\vect{r}\cdot\left(\vect{u} - \frac{\dot{a}}{2a}\vect{r}\right)
  \Bigg\}\Bigg) = \nonumber \\
& & a^{\alpha+2\beta-\nu}\left(H - H_{flow}\right),
\label{Hamiltonian}
\end{eqnarray}
where $H$ is the Hamiltonian of the system in the inertial frame
$\mathbf{X}$ and $\tilde H_{flow}$ is the contribution to the
Hamiltonian due to the global fluid flow. Even though $H$ is
conserved, it follows from the above expression that $\tilde H$ is no
longer invariant under time translation. Moreover, that is the case
even when scaling parameters $\alpha$ and $\beta$ are set to zero to
leave density and pressure fields invariant under the
transformation. Therefore, for the general transformation
$\mathbf{\Lambda}$, in which $a(t)$ and $\vect{\omega(t)}$ are
arbitrary externally set functions of time, energy is no longer a
conserved quantity. Momentum is not conserved due to the forces that
are the consequence of the non-inertiality of the frame. Thus,
non-conservation of momentum and energy, according to Noether's
theorem \citep{Salmon}, does not admit existence of divergence-form
momentum and energy conservation laws. However, it is always possible
to have mass as a conserved quantity by setting $\alpha = \nu$ in
eq. (\ref{rhotr}), as can be seen in eq. (\ref{EulerTr1}).

There exists, however, a subclass of restricted transformations
$\mathbf{\Lambda_R}$ which admits invariance of Euler equations. That
is the subclass of non-rotating reference frames that expand/contract
with a constant velocity, i.e., for which $\dot{a} = Const$. The
corresponding system of equations is obtained by dropping all terms on
the right-hand side of eqs. (\ref{EulerTr1}) - (\ref{EulerTr3}) except
for the first ones\footnote{The terms in curly brackets are indeed
equal to zero due to eq. (\ref{ddot_a}).}. It is immediately clear
that for
\beq
\alpha = \nu, \quad \beta = 1
\label{BestChoice}
\eeq
mass and momentum conservation equations are invariant and, thus, the
new mass and momentum are conserved quantities. The expression inside
the square brackets in eq. (\ref{EulerTr3}) then takes the form
$2\tilde\rho \tilde e - \nu\tilde P$. That suggests that invariance of
the energy conservation equation depends only on the choice of the
equation of state. To illustrate that, introduce independent scaling
of the internal energy $\tilde e = a^{\delta}e$. Then the first law of
thermodynamics in the computational frame becomes
\beq
\frac{D\tilde e}{D\tau} = a^{\delta-2\beta}\frac{\tilde P}{\tilde \rho^2}
\frac{D\tilde \rho}{D\tau} + \left( \delta \tilde \rho \tilde e 
- \alpha a^{\delta - 2\beta}\tilde P\right)\frac{d\ln a}{d\tau}\frac{1}{\tilde \rho}.
\label{FirstLaw}
\eeq
First, in order for the left-hand side of eq. (\ref{EulerTr3}) to have
the traditional divergence form we must set $\delta = 2\beta$, which
justifies the internal energy scaling previously introduced in eq.
(\ref{Ptr}) on the grounds of thermodynamic consistency. Then given
(\ref{BestChoice}) the second term in eq. (\ref{FirstLaw}) is, up to
the factor $\tilde \rho^{-1}$, identical to the right-hand side of
eq. (\ref{EulerTr3}). Hence, as expected, given conservation of new
mass and momentum, in order for energy to be a conserved quantity the
first law of thermodynamics must be invariant under the transformation
$\mathbf{\Lambda}$. Consider the perfect gas equation of state. Then,
eq. (\ref{FirstLaw}) takes the form
\beq
\frac{D\tilde e}{D\tau} = \frac{\tilde P}{\tilde \rho^2}
\frac{D\tilde \rho}{D\tau} + \bigg[ 2\beta -
\alpha(\gamma - 1)\bigg]\frac{d\ln a}{d\tau}\tilde e = 
 \frac{\tilde P}{\tilde \rho^2} \frac{D\tilde \rho}{D\tau} + \tilde T
\frac{d\left(c_v\ln a^{2\beta-\alpha(\gamma -1)}\right)}{d\tau},
\label{FirstLawIdealGas}
\eeq
where $\gamma$ is the polytropic index and $c_v$ is heat capacity at
constant volume. Then combining the requirement for the second term in
the above equation to vanish along with the requirements for the
conservation of new mass and momentum, we obtain the following system
of equations for $\alpha$ and $\beta$
\beq
\left\{ \begin{array}{lcl}
\alpha - \nu \, & = & \, 0, \\
\alpha - \nu + \beta - 1 \, & = & \, 0, \\
2\beta-\alpha(\gamma -1) \, & = & \, 0. \\
\end{array} \right.
\label{ScalingEqs}
\eeq
It is clear that the system is overdetermined and the only solution
admitting the invariance of the fluid equations is indeed
(\ref{BestChoice}) and it exists for the only value of the polytropic
exponent $\gamma = 1 + \frac{2}{\nu}$. In the system (\ref{Euler1}) -
(\ref{Euler3}) there are only three independent physical quantities,
namely length, time, and mass. Thus, having introduced scaling of
length, we have only two degrees of freedom in terms of scaling while
there are three constraints that need to be satisfied in order to
warrant invariance of the original equations.  This fact shows that
fluid equations are not invariant even under the restricted
transformation $\mathbf{\Lambda_R}$ for a general equation of state.
Nevertheless, in the case of expansion/contraction in three dimensions
of gas with $\gamma = \sfrac{5}{3}$, which is extremely important in
astrophysics, fluid equations are invariant under the transformation
$\mathbf{\Lambda_R}$.

The source term in eq. (\ref{EulerTr3}) given the restricted
transformation $\mathbf{\Lambda_R}$ is a ``true'' heating source term
when $\alpha = \nu$ and $\beta = 1$, i.e., it does not depend on the
velocities and acts to change the internal energy of gas. Indeed, the
second term in eq. (\ref{FirstLawIdealGas}), which acts as a heat
source, can be written in the form $\tilde T D\tilde S/D\tau$, where
$\tilde T$ is a modified temperature and $\tilde S = c_v\ln
a^{2\beta-\alpha(\gamma -1)}$ acts as a modified entropy in the
computational frame $\mathbf{\tilde X}$.

The result just obtained based on the general scaling and
thermodynamic arguments has, in fact, a much broader context. The
transformation $\mathbf{\Lambda_R}$, given eq. (\ref{BestChoice}), is
a member of the recently discovered maximal kinematical invariance
group of fluid equations, called the Schr\"odinger group
\citep{Jahn01,Raif,Hassaine}
\beq
\mathcal{G} = GL(2,R)\wedge G,
\eeq
where $G$ is the Galilean transformation subgroup. It can be shown,
that the density field transformation (\ref{rhotr}), and consequently
pressure and internal energy field transformations (\ref{Ptr}) that
follow, are indeed the only transformations that admit invariance of
fluid equations \citep{Raif}. It should also be noted that the
transformation discovered by \citep{Drury} that establishes the
isomorphism between explosion and implosion belongs to such subclass
of restricted transformations $\mathbf{\Lambda_R}$. Moreover, the same
transformation plays an extremely important role in cosmology, e.g.,
it provides conformal mapping of the Kaluza-Klein 5-metric, describing
the relativistic Friedmann universe with constant curvature, to flat
space \citep{Hassaine}.

The great utility of considering such general scaling transformations
of primitive variable fields is a large degree of flexibility they
provide with regards to the source terms, which can be adjusted to the
needs of a particular problem in question. For example, besides the
choice of $\alpha$ and $\beta$ given in eq. (\ref{BestChoice}), the
second important case is
\beq
\alpha = \nu, \quad \beta = \frac{\nu(\gamma - 1)}{2}.
\label{SecondChoice}
\eeq
This provides invariance of the first law of thermodynamics under the
transformation $\mathbf{\Lambda_R}$ for all values of $\gamma$ at the
expense of momentum conservation. However, in this case the source
term in the energy conservation equation is a function only of the
kinetic energy. This may be a preferred choice compared to the one,
discussed before, if thermal energy dominates the local kinetic energy
(while both of those can still be much smaller than the kinetic energy
associated with the global fluid flow).

The key complication associated with the above two choices
(\ref{BestChoice}) and (\ref{SecondChoice}) of scaling parameters
$\alpha$ and $\beta$ is the fact that they modify physical primitive
variable fields. One of the great benefits of eqs. (\ref{EulerTr1}) -
(\ref{EulerTr3}) is the fact that their homogeneous part is
form-invariant compared to the original set of equations. This allows
for quick and straightforward implementation of this method via
operator splitting technique thereby permitting it to be combined
virtually with any available implicit or explicit Eulerian
hydrodynamic scheme and AMR strategy. However, in the case of scaled
primitive variable fields that would require further justification. In
particular, form-invariance of the Rankine-Hugoniot conditions for the
transformed fields as well as the fact that such scaling
transformations result in physical shocks must be shown. That fact has
been recently established by \citep{Jahn05} for the scaling
transformation given by eq. (\ref{BestChoice}) and for the restricted
transformation $\mathbf{\Lambda_R}$. Further generalization of such
proof still remains to be carried out. We refer to \citep{Fazio} for
the discussion of various other aspects of application of scaling
transformations in numerical schemes. Another complication may be due
to the presence of other physical processes in the system. The use of
scaled fields in such source terms may not always be beneficial. One
example are systems governed by a complicated equation state, which
may not be readily adapted to the transformed fields. In situations
when it is desired to avoid the above two complications there exists
the third, most natural, choice of scaling parameters
\beq
\alpha = 0; \quad \beta = 0.
\label{SimplestChoice}
\eeq
In this case the invariance of the original fluid equations is always
broken, however the transformed fields remain physical. Hereafter for
clarity we will designate the transformed time, which corresponds to
such choice of $\beta$, as $\tau'$. This last form of the transformed
equations is the most likely choice for systems involving complex
physics, e.g., SNe explosions. Moreover, it represents the
``worst-case scenario'' in terms of the accuracy of computations as
none of the state vector components are conserved. Hence in the rest
of this paper we will focus on the tests of this particular
formulation also discussing briefly the performance of the method with
the other choices of scaling parameters.

Finally, in the case of the transformation $\mathbf{\Lambda}$ being
dynamical, i.e., in which functions $a(t)$ and $\vect{\omega}(t)$ are
no longer free parameters but instead are uniquely determined by
global distributions of density, pressure, etc., it may be possible to
achieve invariance of fluid equations for a broader class of
transformations than the one discussed above (see also
\citep{GurbatovSaichev}). Moreover, certain systems may admit
asymptotic invariance of the equations. Consider expansion of a gas
sphere into vacuum given ideal gas equation of state with $\gamma =
\sfrac{5}{3}$ and described in an expanding frame $\mathbf{\tilde X}$
with the choice (\ref{BestChoice}) of scaling parameters $\alpha$ and
$\beta$. That test problem is discussed in detail in
\S~\ref{TestTypes}. Initially during the so-called acceleration phase
momentum and energy are not conserved in that system due to the terms
describing reference frame acceleration.  However, eventually the
system asymptotes to the free ballistic expansion, characterized by
expansion at constant velocity. Therefore, asymptotically invariance
of the fluid equations is reached and the conservation of momentum and
energy is achieved.

\section{Numerical Method}
%-------------------------------------------------------------------------------
\label{NumMethod}

We seek to solve eqs. (\ref{EulerTr1}) - (\ref{EulerTr3}) with the
scaling parameters $\alpha$ and $\beta$ given by eq.
(\ref{SimplestChoice}). As it was discussed in
\S~\ref{EulerEquations}, we would like to exploit the form-invariance
of the homogeneous part of fluid equations, thus we employ the
traditional approach of operator splitting.  The method presented in
this work was implemented and tested with two hydrodynamic codes: ALLA
\citep{Khokhlov} and AstroBEAR \citep{Poludnenko}. ALLA code is an AMR
code utilizing the cell-by-cell refinement strategy and, in
particular, the Fully Threaded Tree (FTT) AMR algorithm
\citep{Khokhlov}. The hydrodynamic solver of the code is based on the
dimensionally split scheme that is second-order accurate both in space
and time, in which second-order accuracy in space is achieved via
linear data reconstruction in each cell \citep{Khokhlov}. AstroBEAR
code relies on a different AMR approach, namely the grid-based AMR
\citep{BergerLeVeque}. The solution on each grid is advanced in a
dimensionally unsplit fashion via the second-order accurate wave
propagation scheme \citep{LeVeque}, in which second-order accuracy is
achieved via flux-limiting and proper consideration of transverse wave
propagation. The Riemann problem solution in both codes is obtained
with the exact Riemann solver. As it can be seen, although both codes
are Eulerian, other than that they rely on completely different AMR
strategies and hydrodynamic integration schemes.

We performed testing using both the simplest direct operator splitting,
when the solution $\tilde q^{n+1}$ at the end of the time step $\Delta
t$ is obtained by the successive application of the hydrodynamic
$\mathcal{H}$ and source term $\mathcal{S}$ operators
\beq
\tilde q^{n+1} = \mathcal{S}^{(\Delta t)}\mathcal{H}^{(\Delta t)}\tilde q^n,
\label{OperatorSplitting}
\eeq
as well as Strang splitting
\beq
\tilde q^{n+1} = \mathcal{S}^{(\sfrac{\Delta t}{2})}\mathcal{H}^{(\Delta t)}
\mathcal{S}^{(\sfrac{\Delta t}{2})}\tilde q^n.
\label{StrangSplitting}
\eeq
In the above $\mathcal{H}$ represents the left-hand side of
eqs.(\ref{EulerTr1}) - (\ref{EulerTr3}), while $\mathcal{S}$
represents their right-hand side. Strang splitting can require
significantly larger computational effort than the direct operator
splitting approach, especially in the case of implicit source term
solvers, which can be an important consideration in large-scale
three-dimensional simulations. Therefore, when studying the accuracy
and conservativity properties of the method presented here we
primarily focus on the first approach to obtain the upper bounds on
the accuracy and conservativity errors, though we also discuss the
effect that the use of Strang splitting has on those errors. On the
other hand, direct operator splitting is formally only first-order
accurate, hence we employ Strang splitting in convergence studies to
demonstrate that it is possible to achieve second-order accuracy with
our method.

The AMR kernels and hydrodynamic solvers of the ALLA and AstroBEAR
codes were not modified from their original form. The CFL condition
in the moving reference frame $\mathbf{\tilde X}$ retains its usual form
thus limiting the computational time step $d\tau$ as
\beq
d\tau \leq \min_i\left\{\frac{d\tilde x}{\tilde u_i + \tilde
c_i}\right\}.
\label{CFLmovingframe}
\eeq
Here the spatial step in the frame $\mathbf{\tilde X}$ is $d\tilde x =
dx/a$, the transformed sound speed is $\tilde c = a^{\beta} c$, and
the velocity field $\tilde u$ is defined by eq. (\ref{newvel}).

A single source term integrator was implemented and used in both
codes. The integrator was developed as a standalone module which, with
an appropriate data wrapper, could be used with any Eulerian
hydrodynamic code. Performance of the presented method depends very
sensitively on the quality of the source term solver. Source terms in
eqs.  (\ref{EulerTr1}) - (\ref{EulerTr3}) can be very stiff as grid
accelerations can be quite large. Consequently, explicit source term
solvers can be either completely unacceptable or their use may lead to
significantly shorter time steps and much more inferior solution
accuracy. Thus, we chose to use the 4th-order accurate implicit
Rosenbrock method, in particular its implementation by Kaps and
Rentrop \citep{KR79,PTVF97}. This method for moderate accuracies
($\epsilon \lesssim 10^{-4} - 10^{-5}$ in relative error) and
modest-sized systems, such as eqs.  (\ref{EulerTr1}) -
(\ref{EulerTr3}), is competitive with, yet simpler than, more
complicated algorithms, e.g., semi-implicit extrapolation method
\citep{PTVF97}. It is the lowest order implicit scheme that is
embedded, i.e., which provides error control and adaptive stepsize
adjustment. This feature not only permits explicit monitoring of the
solution accuracy but it also allows fine-tuning of the solver to
achieve the desired balance between the performance and the acceptable
error level. In particular, as it will be discussed in
\S~\ref{Conservativity}, this gives the means to control the
conservativity properties of the solution. The implemented solver is
capable of integrating arbitrary systems of source terms that are
functions only of temporal and spatial coordinates. In particular, if
there are other source terms in the problem in question, e.g.,
geometric, gravity, energy release source terms, etc., as often is the
case in complex multi-physics simulations, the computation in a moving
frame can be performed at virtually no, or minimal, extra
computational cost. For a specific choice of source terms one simply
must provide their description as well as the Jacobian matrix based on
the source term functions $f_i$\footnote{This should not be confused
with the Jacobian of the flux functions of eqs. (\ref{EulerTr1}) -
(\ref{EulerTr3}).}  ${A_{ij}}(\tilde q_j)=\partial f_i/\partial \tilde
q_j$, where $\tilde q_j$ are the state vector components. The explicit
expression for the Jacobian matrix is rather cumbersome and we will
not show it here. 

Two important points must be emphasized. Firstly, in the case of stiff
source terms the adaptive stepsize control will lead to time step
subcycling in the source term integration over the hydrodynamic time
step. This is done in order to maintain the solver accuracy, in
particular the solver ensures that the desired relative error has been
achieved and that the solution during the current substep has not
changed by more than a certain percentage. Secondly, in the presence
of extremely strong source terms the method described above can also
fail producing negative pressures. In order to prevent this we
included adaptivity in time in the implicit integration. Source term
functions in eqs. (\ref{EulerTr1}) - (\ref{EulerTr3}) as well as the
Jacobian ${A_{ij}}$ carry explicit dependence on time due to the
presence of terms that contain temporal derivatives of $\ln a$ and
$\ln \Omega$ (see also expressions for the latter in
\S~\ref{FrameTypes}). However, we find that it is highly beneficial
for the accuracy and stability of the solution to assume that source
term functions and the Jacobian do not depend on time for all
subcycling time steps and to use the time value that corresponds to
the beginning of the global hydrodynamic time step. This also applies
to both substeps in the Strang splitting approach
(\ref{StrangSplitting}). This issue will be discussed in further detail
in \S~\ref{Accuracy}.

Initial conditions in the computational frame are obtained by applying
the transformation $\mathbf{\Lambda}$ as well as the velocity
transformation (\ref{newvel}) and density and pressure field
transformations (\ref{rhotr}) and (\ref{Ptr}) to the initial
conditions in the physical space. It is convenient to set the initial
value of the scale factor $a(t=0) = 1$. This ensures that at $t=0$ we
have $\vect{\tilde r} = \vect{r}$ and $\tau = 0$, i.e., physical and
computational coordinate systems initially coincide. The choice of the
initial expansion/contraction rate $\dot{a}$ and, if necessary,
initial angular velocity of the grid is dictated by the problem
itself. We give examples of that in the discussion of test problems
below.

There may be several different possibilities for the specification of
boundary conditions in a computational domain advanced in the
reference frame $\mathbf{\tilde X}$\footnote{Note, that here we
consider only the outer boundaries with respect to the fixed point of
expansion/contraction.  The boundaries that contain the fixed point
itself are typically set to be perfectly reflective.}. The first
possibility is the case when the expanding flow is fully contained
within the computational domain and the ambient material is
dynamically unimportant, e.g., if it represents ``numerical
vacuum''. In this case the best strategy is to set ambient material at
zero velocity in the computational frame and then, for all reference
frame types other than constant velocity expanding/contracting frames
which automatically maintain that zero velocity, keep it at that value
throughout the simulation. This prevents ambient material from
developing significant velocities as the reference frame
accelerates. Then boundary conditions can be set to be either
perfectly reflective or zero-order extrapolation (outflow). This
approach, employing perfectly reflective boundary conditions, was used
in tests involving expansion of a non-rotating sphere into vacuum
discussed below. The second possibility is the case when the
computational domain contains only the central region of the expanding
or contracting flow, i.e., the global flow crosses the outer
boundaries of the domain. In this case the flow in the computational
frame would be transonic or subsonic. Consequently, the use of the
standard zero-order extrapolation (outflow) boundary conditions can
lead to the formation of spurious features propagating from the
boundary. Thus, boundary conditions better suited for subsonic flows,
e.g., characteristic boundary conditions, may be required. We use
simpler zero-order extrapolation boundary conditions in tests
involving the converging shock wave since, as it will be discussed in
\S~\ref{TestTypes}, the solution in the vicinity of the shock front which
we are primarily interested in is insensitive to the boundary
conditions. Finally, the third possibility is the case least suited
for the method presented here, as mentioned in \S~\ref{GeneralFormalism},
i.e., the case when the ambient material is dynamically important and,
therefore, its velocity cannot be adjusted to that of the reference
frame. The most immediate example is the ambient material stationary
in the inertial reference frame. Then, in the computational frame that
material will have velocity $\vect{\tilde u} = -\dot{a}
\vect{\tilde r}$, which can be quite large. Consequently, two
approaches can be adopted in this case. In the first approach at the
end of a time step the values of $a$ and $\dot{a}$ are determined for
the next time step based either on the a priori analytic prescription
or on the fluid motion itself. Then ghost cells are initialized by
setting density and pressure to their specified ambient values, while
setting the velocity in ghost cells to $\tilde u_i = -\dot{a}\tilde
x_i$, where $\tilde x_i$ are the ghost cell center coordinates. This
``inflow'' represents the stationary ambient material engulfed by the
expanding computational domain. We take this approach in setting
boundary conditions in tests involving strong point explosion. While
we find this method to be exceptionally simple, as it does not require
any use of the interior cells of the domain, and yet accurate, it
still may result in small noise-like features propagating away from
the boundaries. This happens when there is a slight mismatch between
the expected values of $a$ and $\dot{a}$ used to set ghost cells, and
the ones that are based on the actual linear velocity profile in the
regions of stationary ambient material. Such linear velocity profile
may deviate due to numerical errors from the correct one which would
correspond to the material stationary in the inertial frame. Using the
expected values of $a$ and $\dot{a}$ to set ghost cells can lead to a
break of the linear velocity profile at the boundary and, thus, cause
the formation of unphysical waves propagating from the boundaries. To
avoid this situation the ghost cell density and pressure can be set
based on the values obtained from the adjacent interior
cells\footnote{A simpler approach would be to set density and
pressure, as in the previous case, to their specified ambient
values. However, as with $\dot{a}$, numerical errors can lead to
slight discrepancies between the actual and pre-defined values of
ambient density and pressure which, in their turn, can lead to the
formation of waves propagating away from the boundaries.} while the
velocity is set as $\tilde u_i = \left(\tilde u_i^*/\tilde
x_i^*\right)\tilde x_i$. Here $\tilde u_i^*$ and $\tilde x_i^*$ refer
to the interior cell nearest to the current ghost cell. Note, that
this still leaves the ambiguity in specifying the corner ghost
cells. These cells are set by applying the above procedure to the
nearest non-corner ghost cells, that have already been set this
way. In our experience the above approach, while being a bit more
complicated in implementation and somewhat more taxing in terms of
runtime overhead, eliminates any features that may propagate from the
boundaries and, therefore, can be used when highly noise-free boundary
conditions are required.

Addition of rotation in the first case considered above, i.e., the
case in which ambient material is dynamically unimportant, does not
change the situation. Again the best strategy is to set ambient
material velocity to zero and maintain it at that value throughout the
simulation. We use this approach in combination with zero-order
extrapolation boundary conditions in the tests involving expansion of
a rotating sphere into vacuum discussed below. In all other cases
addition of rotation may significantly complicate matters, however, we
leave that discussion outside the scope of this work as we do not
utilize other boundary condition types in tests discussed below.

\section{Numerical Tests}
%-------------------------------------------------------------------------------
\label{NumTests}

\subsection{Types of considered computational reference frames}
%-------------------------------------------------------------------------------
\label{FrameTypes}

In all numerical tests we use five types of computational reference
frames that are discussed below. In all cases we assume the reference
frame origin to be located at the point $x_i = 0$ which is the fixed
point of expansion/contraction. In order to define a particular
reference frame one has to specify three scales:

(1) \emph{length scale}, which is typically the extent $\tilde r_d$ of
the computational domain (details regarding the specification of the
domain extent will be given in the discussion of the setup of individual
tests);

(2) time scale, which is the total physical run time $t_{tot}$ of the
simulation; and 

(3) velocity scale, or the velocity $v_g = \dot{a} \tilde r_d$ of the
computational grid (detailed meaning of this parameter for each type
of reference frame will be given below).

As it was discussed in \S~\ref{GeneralFormalism}, we primarily focus
on the choice of scaling parameters (\ref{SimplestChoice}). In order
to specify the transformation $\mathbf{\Lambda}$ as well as the
primitive variable field transformations (\ref{newvel}),
(\ref{rhotr}), and (\ref{Ptr}) we need to provide the description of
the scale factor $a(t)$, angular velocity $\omega(t)$ of the frame
$\mathbf{\tilde X}$, and the resulting temporal transformation of the
physical time $t$ to the computational time $\tau'$. This allows one
to determine the total computational run time of the simulation based
on the desired total physical run time. We also need to provide the
inverse temporal transformation $\tau' \to t$. All this allows one
then to obtain temporal derivatives of $\ln a$ and $\ln \Omega$ that
can be substituted into the eqs. (\ref{EulerTr1}) - (\ref{EulerTr3})
to obtain the set of fluid equations, transformed to the computational
frame, that is being solved. One can then also use expressions for
$a(t)$ and $\dot{a}(t)$ as well as the temporal transformation $\tau'
\to t$ in order to perform the remap of the computational domain back
into the physical frame.

\emph{Type a. Constant velocity expanding/contracting reference frame.}

In this case we set $a(t=0) = 1$, $\dot{a} = Const$, and
$\vect{\omega} = 0$. Grid velocity $v_g = \dot{a} \tilde r_d$ is the
velocity of the coordinate $\tilde r_d$ corresponding to the edge of
the computational domain with respect to the inertial frame
$\mathbf{X}$. Then
\beq
\left\{ \begin{array}{lcl}
a(t)    & = & \displaystyle \frac{v_g t}{\tilde r_d} + 1, \\
\dot a  & = & \displaystyle \frac{v_g}{\tilde r_d}, \\
\ddot a & = & 0.
\end{array} \right.
\label{adotConst}
\eeq
Note that $v_g$ can be both positive (expansion) and negative
(contraction), with the only restriction that $a(t) > 0$ always, i.e.,
in the case of contraction
\beq
t < -\frac{\tilde r_d}{v_g}.
\label{taConstraint}
\eeq
Then direct and inverse temporal transformations in this case are
\begin{eqnarray}
\tau' & \, = \, & \frac{\tilde r_d}{v_g}\ln \left( \frac{v_g t}{\tilde r_d}
+ 1 \right) = \frac{\tilde r_d}{v_g}\ln a,
\label{tau_a} \\
t     & \, = \, & \displaystyle \frac{\tilde r_d}{v_g}
\left( \e^{\frac{v_g \tau'}{\tilde r_d}} - 1\right).
\label{t_a}
\end{eqnarray}
Finally, it follows from eq. (\ref{tau_a}) that
\begin{eqnarray}
\pd{\ln a}{\tau'}  & \, = \, & \frac{v_g}{\tilde r_d}, 
\label{dlna_a}\\
\pdd{\ln a}{\tau'} & \, = \, & 0.
\label{ddlna_a}
\end{eqnarray}

We note that we also use a variation of this reference frame type,
namely a constant velocity expanding/contracting frame with the
delayed stretch, that we designate as \emph{type $a_d$}. In this case
the computational domain is initially advanced in the inertial frame
$\mathbf{X}$. Then at a certain moment in time $t_s$ the computational
domain is transformed from the frame $\mathbf{X}$ to the computational
frame $\mathbf{\tilde X}$ noting that in all of the above expressions
from that moment on $t = t_{run} - t_s$, where $t_{run}$ is the
physical time elapsed since the start of the simulation. Then at $t =
t_s$: $\vect{\tilde r} = \vect{r}$, $\tau' = 0$, and $\vect{\tilde
u}(\vect{\tilde r}) = \vect{u}(\vect{r}) -
\dot{a}\vect{r}$.

\emph{Type b. Constant acceleration expanding reference frame.}

In this case we set $a(t=0) = 1$, $\dot a(t=0) = 0$, $\ddot a =
Const$, and $\vect{\omega} = 0$. Grid velocity $v_g$ in this case is
the velocity of the edge of the computational domain $\tilde r_d$ at the end
of the simulation
\beq
v_g = \dot a \tilde r_d = \ddot a t_{tot} \tilde r_d,
\eeq
where $t_{tot}$ is the total physical run time of the simulation. Then
\beq
\left \{ \begin{array}{lcl}
a(t)      & = & \displaystyle \ddot a \frac{t^2}{2} + 1, \\ 
\dot a(t) & = & \ddot a t, \\
\ddot a   & = & \displaystyle \frac{v_g}{\tilde r_d t_{tot}}.
\end{array} \right.
\label{addotConst}
\eeq
Temporal transformations then take the form
\begin{eqnarray}
\tau' & \, = \, & \sqrt{\frac{2}{\ddot a}}\tan^{-1}
\left( \sqrt{\frac{\ddot a}{2}}t \right),
\label{tau_b} \\
t    & \, = \, & \sqrt{\frac{2}{\ddot a}}\tan
\left( \sqrt{\frac{\ddot a}{2}}\tau' \right).
\label{t_b}
\end{eqnarray}
Substituting eq. (\ref{t_b}) into the expressions for $\dot a(t)$ and
$a(t)$ (eq. (\ref{addotConst})) and recalling eq. (\ref{dot_a}) and
(\ref{ddot_a}) we get
\begin{eqnarray}
\pd{\ln a}{\tau'}  & \, = \, & \sqrt{2\ddot a}\tan
\left( \sqrt{\frac{\ddot a}{2}}\tau' \right),
\label{dlna_b} \\
\pdd{\ln a}{\tau'} & \, = \, & \ddot a
\left( \tan^2\sqrt{\frac{\ddot a}{2}}\tau' + 1\right).
\label{ddlna_b}
\end{eqnarray}

\emph{Type c. Constant acceleration contracting reference frame.}

The principal difference of this case from the previous one is that
$v_g < 0$. Since $a(t) > 0$ always, the following condition follows
from the expression for $a(t)$ in eq. (\ref{addotConst})
\beq
t < -\frac{2\tilde r_d}{v_g}.
\label{tcConstraint}
\eeq
While expressions for $a(t)$ and its temporal derivatives are the same
as in eq. (\ref{addotConst}), the computational time $\tau'$ and the
corresponding inverse temporal transformation are obtained by
performing the integration in eq. (\ref{Transform}) while taking
proper account of the above constraint and the fact that $v_g < 0$
\begin{eqnarray}
\tau' & \, = \, & \displaystyle \frac{1}{\sqrt{-2\ddot a}}
\ln \frac{1 + \sqrt{-\frac{\ddot a}{2}}t} {1 - \sqrt{-\frac{\ddot a}{2}}t},
\label{tau_c} \\
t    & \, = \, & \displaystyle \sqrt{-\frac{2}{\ddot a}} \,
\frac{\e^{\sqrt{-2\ddot a}\tau'} - 1}{\e^{\sqrt{-2\ddot a}\tau'} + 1},
\label{t_c}
\end{eqnarray}
where $\ddot a$ is again defined in eq. (\ref{addotConst}). Finally,
substituting eq. (\ref{t_c}) into the expression for $\dot a(t)$
(eq. (\ref{addotConst})) we obtain
\beq
\pd{\ln a}{\tau'} = -\sqrt{-2\ddot a} \,
\frac{\e^{\sqrt{-2\ddot a}\tau'} - 1}{\e^{\sqrt{-2\ddot a}\tau'} + 1},
\label{dlna_c}
\eeq
while substituting eq. (\ref{t_c}) into the expression for $a(t)$
(eq. (\ref{addotConst})) and then making use of eq. (\ref{ddot_a}) we
get
\beq
\pdd{\ln a}{\tau'} = \ddot a \left( 1 - \left\{
\frac{\e^{\sqrt{-2\ddot a}\tau'} - 1}{\e^{\sqrt{-2\ddot a}\tau'} + 1}
\right\}^2 \right).
\label{ddlna_c}
\eeq

\emph{Type d. Oscillating reference frame.}

We define the noninertial reference frame $\mathbf{\tilde X} =
\{\tilde \vect{r},\tau'\}$ that oscillates with respect to the inertial
frame $\mathbf{X}$ in a sinusoidal fashion. The transformation
$\mathbf{\Lambda}$ in this case is subject to the conditions $a(t=0) =
1$, $\dot a(t=0) = 0$, $\ddot a(t=0) = \ddot a_0$, and $\vect{\omega}
= 0$. Then
\beq
\left \{ \begin{array}{lcl}
a(t)       & = & \displaystyle \ddot a_0\varphi^2
                 \left( 1 - \cos\frac{t}{\varphi} \right) + 1, \\ 
\dot a(t)  & = & \displaystyle \ddot a_0\varphi\sin\frac{t}{\varphi}, \\
\ddot a(t) & = & \displaystyle \ddot a_0 \cos\frac{t}{\varphi},
\end{array} \right.
\label{adotOscil}
\eeq
where $\varphi = t_p/2\pi$ and $t_p$ is the duration of one period of
oscillation in physical time. In this case grid velocity $v_g = \dot
a(t=t_p/4) \tilde r_d$ is the maximum velocity of the edge of the
computational domain in the course of one oscillation period, i.e., at
the time $t = t_p/4$. Using this in the expression for $\dot a(t)$
above we find
\beq
\ddot a_0 = \frac{2\pi v_g}{t_p \tilde r_d}.
\eeq
Performing integration in eq. (\ref{Transform}) using expression for
$a(t)$ from eq. (\ref{adotOscil}) we obtain the direct and inverse
temporal transformations
\begin{eqnarray}
\tau' \, & = \, & \frac{2\varphi}{\sqrt{2\ddot a_0 \varphi^2 + 1}}\tan^{-1}\left\{
\sqrt{2\ddot a_0 \varphi^2 + 1}\tan\frac{t}{2\varphi}\right\},
\label{tau_d} \\
t     \, & = \, & 2\varphi\tan^{-1}\left\{ \frac{1}{\sqrt{2\ddot a_0 \varphi^2 + 1}}
\tan\left( \frac{\sqrt{2\ddot a_0 \varphi^2 + 1}}{2\varphi}\tau'\right)\right\}.
\label{t_d}
\end{eqnarray}
Finally, using eqs. (\ref{adotOscil}) and (\ref{t_d}) in
eqs. (\ref{dot_a}) and (\ref{ddot_a}) we find
\begin{eqnarray}
\pd{\ln a}{\tau'}  & \, = \, & \displaystyle \ddot a_0\varphi\sin\frac{t}{\varphi},
\label{dlna_d} \\
\pdd{\ln a}{\tau'} & \, = \, & \displaystyle \ddot a_0 \cos\frac{t}{\varphi} \left\{
\displaystyle \ddot a_0\varphi^2 \left( 1 - \cos\frac{t}{\varphi} \right) + 1 \right\}.
\label{ddlna_d}
\end{eqnarray}

\emph{Type e. Constant velocity expanding/contracting and rotating reference frame.}

We consider the expanding/contracting and rotating reference frame
$\mathbf{\tilde X}$, that is defined by the transformation
$\mathbf{\Lambda}$. We assume constant velocity of
expansion/contraction. Then expressions for $a(t)$, its temporal
derivatives, and physical and computational times $t$ and $\tau'$ are
the same as in the case of the \emph{type a} reference frame above
(eqs. (\ref{adotConst}) - (\ref{ddlna_a})).

While the choice of a specific expression for angular velocity
$\vect{\omega}(t)$ of the rotating frame $\mathbf{\tilde X}$ is
problem-specific, we point out two special cases that already
encompass a large class of applications.

\emph{1. Constant angular velocity}

Here we assume that $\vect{\omega} = \vect{\omega}_0 = Const$.  This
most closely corresponds to the situations in which the fluid material
is rotating and exhibits no, or very small degree of, global expansion
or contraction, such as in gravitationally bound systems, e.g.,
rotating white dwarfs. Then effective angular velocity is
$\vect{\Omega} = a\vect{\omega} = a\vect{\omega}_0$ and
\beq
\frac{d\ln \Omega}{d\tau'} = \frac{d\ln a}{d\tau'}.
\label{dlnOmega_e1}
\eeq

\emph{2. Expansion-correlated angular velocity}

In systems, that expand (contract) significantly on their dynamical
timescale, conservation of angular momentum causes the fluid to lose
(gain) angular velocity very rapidly. In the noninertial frame
$\mathbf{\tilde X}$ that is initially corotating with the fluid and
whose angular velocity with respect to the inertial frame $\mathbf{X}$
is held constant, such rapid loss (gain) of angular velocity in the
frame $\mathbf{X}$ by the expanding (contracting) material results in
it developing a significant rotational component in the frame
$\mathbf{\tilde X}$. This can render the whole method ineffective. A
better approach would be to have the frame $\mathbf{\tilde X}$
decrease its angular velocity in a manner correlated with its
expansion (contraction) which, in its turn, is governed by fluid
motions. For this $\vect{\omega}(t)$ can be found via the following
simple argument. Consider a region that initially extends from 0 to
$r_0$ and contains fluid of uniform density $\rho_0$ rotating with
angular velocity $\vect{\omega_0}$ in the frame $\mathbf{X}$. Assume
that this region expands (contracts) with the scale factor
$a(t)$. Density of the fluid in that region will then change as
$\rho(t)=\rho_0 (r_0/r_1)^\nu = \rho_0/a^\nu$, where $r_1 = ar_0$.
Then in two dimensions conservation of total angular momentum of the
whole region gives
\beq
\displaystyle M = \int_0^{r_0} 2\pi r\rho_0\vect{\omega}_0r^2dr = 
\int_0^{r_1} 2\pi r\rho\vect{\omega} r^2dr.
\eeq
Substituting expression for $\rho(t)$ in the above equation, finding the
integrals, and solving for $\vect{\omega}$ we find
\beq
\vect{\omega}(t) = \frac{\omega_0}{a^2}.
\label{omega_e2d}
\eeq
In three dimensions angular momentum conservation gives
\beq
\displaystyle M = \int_0^{\pi}\int_0^{r_0} 
2\pi \rho_0\vect{\omega}_0 r^3\sin^3\theta drd\theta =
\int_0^{\pi}\int_0^{r_1} 2\pi \rho\vect{\omega} r^3\sin^3\theta
drd\theta.
\eeq
Again, via the same steps as before we find
\beq
\vect{\omega}(t) = \frac{\omega_0}{a}.
\label{omega_e3d}
\eeq
Consequently
\beq
\frac{d\ln\Omega}{d\tau'} = \left\{ \begin{array}{ll}
\displaystyle -\frac{d\ln a}{d\tau'} & \quad \textrm{in 2D,} \\
              0                     & \quad \textrm{in 3D.}
\end{array} \right.
\label{dlnOmega_e2} 
\eeq

\subsection{Types of tests}
%-------------------------------------------------------------------------------
\label{TestTypes}

\begin{deluxetable}{ccccccccc}
\tablecaption{Summary of the Runs Discussed \label{Runs1}}
\tablenum{1a}
\tabletypesize{\small}
\tablewidth{0pt}
\tablehead{
\colhead{}                             &
\colhead{Test}                         &
\colhead{$N_{dim}$}                    &
\colhead{Frame type\tablenotemark{a}}  &
\colhead{Resolution}                   &
\colhead{$v_g$}                        &
\colhead{$N_{osc}$\tablenotemark{b}}   &
\colhead{$\gamma$} }
\startdata
1   &  Sedov 	      &  2D  &  S              &  64 - 512    &  -        &  -    &  $\sfrac{7}{5}$  \\
2   &  Sedov 	      &  2D  &  a              &  64 - 512    &  20.0     &  -    &  $\sfrac{7}{5}$  \\
3   &  Sedov	      &  2D  &  b              &  64 - 512    &  100.0    &  -    &  $\sfrac{7}{5}$  \\ 
4   &  Sedov          &  2D  &  d              &  64 - 512    &  20.0     &  1    &  $\sfrac{7}{5}$  \\
5   &  Sedov          &  2D  &  d              &  64 - 512    &  2.0      &  100  &  $\sfrac{7}{5}$  \\ 
6   &  Sedov          &  3D  &  S              &  256         &  -        &  -    &  $\sfrac{7}{5}$  \\
7   &  Sedov          &  3D  &  b              &  256         &  100.0    &  -    &  $\sfrac{7}{5}$  \\
8   &  Guderley       &  2D  &  S              &  256         &  -        &  -    &  $\sfrac{7}{5}$  \\
9   &  Guderley       &  2D  &  c              &  256 - 512   &  -1000.0  &  -    &  $\sfrac{7}{5}$  \\
10  &  GuderleyShell  &  2D  &  S              &  4096        &  -        &  -    &  $\sfrac{7}{5}$  \\
11  &  GuderleyShell  &  2D  &  c              &  4096        &  -1250.0  &  -    &  $\sfrac{7}{5}$  \\
12  &  Sphere         &  2D  &  S              &  256 - 4096  &  -        &  -    &  $\sfrac{5}{3}$  \\
13  &  Sphere         &  2D  &  a              &  128 - 2048  &  50.0     &  -    &  $\sfrac{5}{3}$  \\
14  &  Sphere         &  2D  &  $\textrm{a}_d$ &  256 - 4096  &  42.5     &  -    &  $\sfrac{5}{3}$  \\
15  &  SphereRot      &  2D  &  S              &  512 - 4096  &  -        &  -    &  $\sfrac{5}{3}$  \\
16  &  SphereRot      &  2D  &  $\textrm{e}_2$ &  256 - 2048  &  80.0     &  -    &  $\sfrac{5}{3}$  \\
17  &  Clump          &  2D  &  a              &  64  - 1024  &  100.0    &  -    &  $\sfrac{5}{3}$  \\
\tablenotetext{a}{ Reference frame type as discussed in \S~\ref{FrameTypes}.
Tests performed in the stationary frame $\mathbf{X}$ are designated
with the letter $\mathbf{S}$. Frame type $e_2$ is the constant
velocity expanding frame with expansion-correlated angular velocity.}
\tablenotetext{b}{ Number of reference frame oscillations in the 
course of a simulation. Note that the duration of one oscillation 
period, used in eq. (\ref{adotOscil}), is $t_p = t_{tot}/N_{osc}$.}
\enddata
\end{deluxetable}

\begin{deluxetable}{ccccccc}
\tablecaption{Summary of the Runs Discussed \label{Runs2}}
\tablenum{1b}
\tabletypesize{\small}
\tablewidth{0pt}
\tablehead{
\colhead{}                                      &
\colhead{$\omega_s$\tablenotemark{a}}           &
\colhead{$\omega$\tablenotemark{b}}             &
\colhead{Domain, $t_{start}$\tablenotemark{c}}  &
\colhead{Domain, $t_{end}$\tablenotemark{c}}    &
\colhead{$t_{tot}$\tablenotemark{d}}            &
\colhead{$\tau'_{tot}$\tablenotemark{d}} }
\startdata
1  &  -      &  -      &  0.0 - 0.3      &  0.0 - 0.3     &  $0.9975\cdot10^{-3}$  &  $0.9975\cdot10^{-3}$ \\
2  &  -      &  -      &  0.0 - 0.28005  &  0.0 - 0.3     &  $0.9975\cdot10^{-3}$  &  $0.9636\cdot10^{-3}$ \\
3  &  -      &  -      &  0.0 - 0.250125 &  0.0 - 0.3     &  $0.9975\cdot10^{-3}$  &  $0.9382\cdot10^{-3}$ \\
4  &  -      &  -      &  0.0 - 0.3      &  0.0 - 0.3     &  $0.9975\cdot10^{-3}$  &  $0.9975\cdot10^{-3}$ \\
5  &  -      &  -      &  0.0 - 0.3      &  0.0 - 0.3     &  $0.9975\cdot10^{-3}$  &  $0.9975\cdot10^{-3}$ \\
6  &  -      &  -      &  0.0 - 0.3      &  0.0 - 0.3     &  $0.355\cdot10^{-3}$   &  $0.355\cdot10^{-3}$  \\
7  &  -      &  -      &  0.0 - 0.28225  &  0.0 - 0.3     &  $0.355\cdot10^{-3}$   &  $0.3478\cdot10^{-3}$ \\
8  &  -      &  -      &  0.0 - 1.0      &  0.0 - 1.0     &  $0.1\cdot10^{-2}$     &  $0.1\cdot10^{-2}$    \\
9  &  -      &  -      &  0.0 - 1.0      &  0.0 - 0.5     &  $0.1\cdot10^{-2}$     &  $0.1246\cdot10^{-2}$ \\
10 &  -      &  -      &  0.0 - 1.0      &  0.0 - 1.0     &  $0.7129\cdot10^{-3}$  &  $0.7129\cdot10^{-3}$ \\
11 &  -      &  -      &  0.0 - 1.0      &  0.0 - 0.55442 &  $0.7129\cdot10^{-3}$  &  $0.8611\cdot10^{-3}$ \\
12 &  -      &  -      &  0.0 - 1.2      &  0.0 - 1.2     &  $0.24\cdot10^{-1}$    &  $0.24\cdot10^{-1}$   \\
13 &  -      &  -      &  0.0 - 0.6      &  0.0 - 38.1    &  $0.75$                &  $0.4981\cdot10^{-1}$ \\
14 &  -      &  -      &  0.0 - 1.2      &  0.0 - 32.2    &  $0.75$                &  $0.1138$             \\
15 &  100.0  &  0.0    &  -1.2 - 1.2     &  -1.2 - 1.2    &  $0.2\cdot10^{-1}$     &  $0.2\cdot10^{-1}$    \\
16 &  100.0  &  100.0  &  -0.6 - 0.6     &  -60.6 - 60.6  &  $0.75$                &  $0.3461\cdot10^{-1}$ \\
17 &  -      &  -      &  0.0 - 0.6      &  0.0 - 75.6    &  $0.75$                &  $0.2902\cdot10^{-1}$ \\
\tablenotetext{a}{ Initial angular velocity of the sphere.}
\tablenotetext{b}{ Initial angular velocity of the computational
reference frame $\mathbf{\tilde X}$.}
\tablenotetext{c}{ Domain extent in physical space at the start
and the end of each simulation.Note that the initial domain extent in
physical space defines the extent of the domain in computational space
throughout the simulation.}
\tablenotetext{d}{ Total physical and computational time of each simulation.}
\enddata
\end{deluxetable}

Tables~\ref{Runs1} and \ref{Runs2} list all numerical tests discussed
in this paper as well as all key parameters describing each test.
Naming convention for designating each test in this work is as
follows: the name of each test is comprised of the values in the first
six columns of Table~\ref{Runs1}\footnote{We do not include the test
number indicated in Table~\ref{Runs1}.}. For example, test $\#5$ is
designated as ``Sedov.2D.d.256.2.100''. All tests used the ideal gas
equation of state and the last column ``$\gamma$'' in
Table~\ref{Runs1} shows the polytropic index value in each
simulation. All tests were performed with the ALLA code except for the
tests ``Guderley'', which were carried out with AstroBEAR. In all
tests the CFL number was 0.7 and the computational domain size
was the same in all dimensions.

\emph{1. Strong point explosion (Sedov blast wave)}

\begin{figure}
\epsscale{0.55}
\plotone{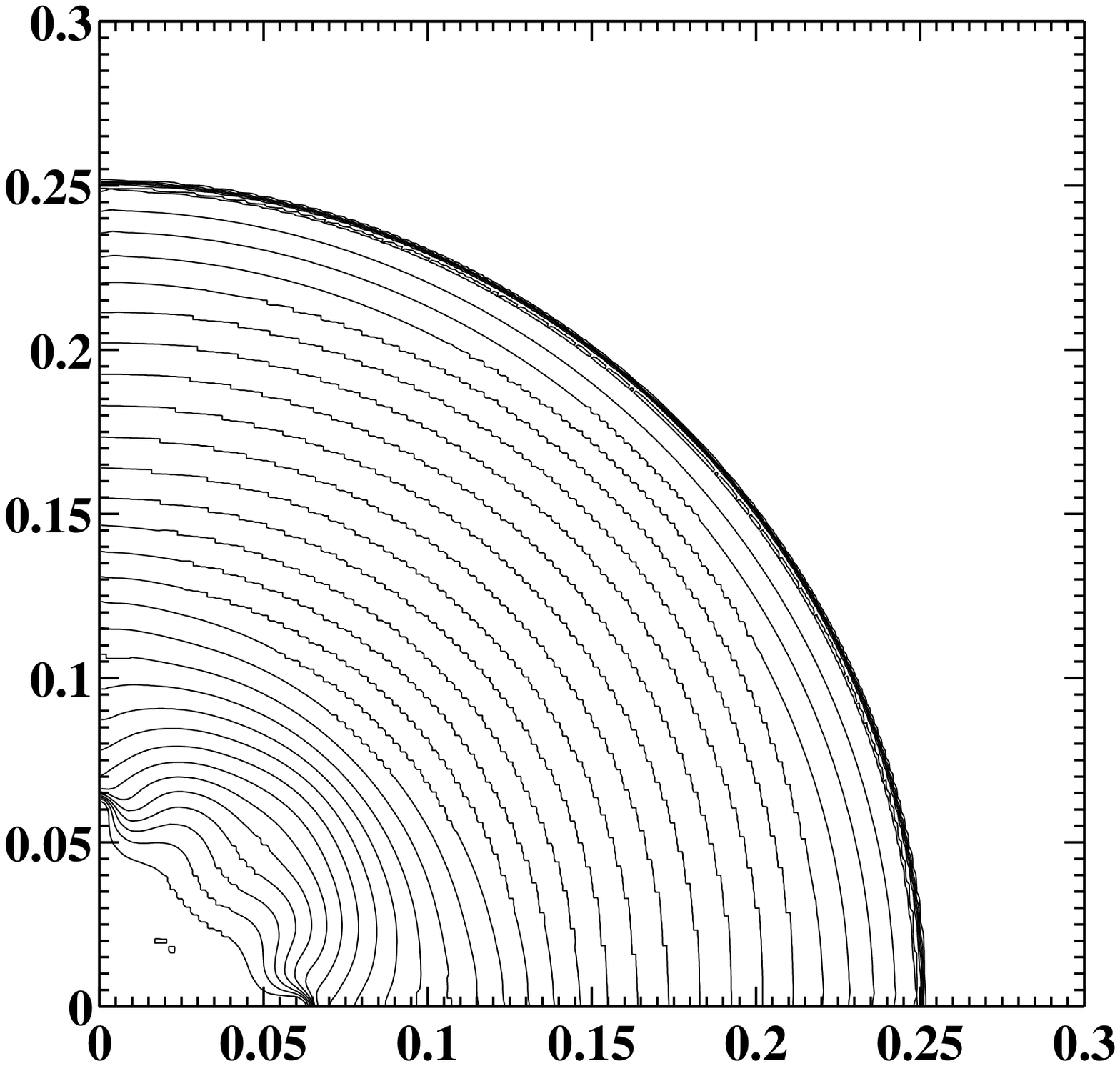}
\caption{Contour plot of the density logarithm for the run Sedov.2D.b.256.100
at time $t_{tot}=0.9975\cdot 10^{-3}$ (cf. Fig.~\ref{SedovCut}). 30
contour lines correspond to the values of the density logarithm that
are equally spaced between its maximum (0.75) and minimum (-3.11)
values.
\label{Sedov2DImage}}
\end{figure}

\begin{figure}[t]
\epsscale{1.03}
\plottwo{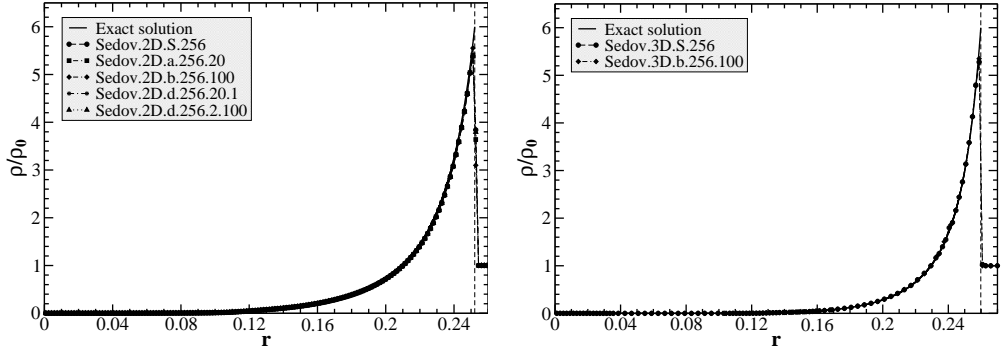}{fig2b.eps}
\caption{\emph{Left:} Comparison of runs Sedov.2D at time
$t_{tot}=0.9975\cdot 10^{-3}$. \emph{Right:} Comparison of runs
Sedov.3D at time $t_{tot}=0.355\cdot 10^{-3}$. In both panels shown is
density along the diagonal cut of the computational domain in the
units of ambient density $\rho_0$. Vertical dashed line indicates the
shock front position based on the exact solution.
\label{SedovCut}}
\end{figure}

In this type of tests, designated as ``Sedov'' in Table~\ref{Runs1},
the computational domain is initialized with a small uniform region of
very high pressure. The size of this region, or charge, is one cell at
the finest refinement level. High pressure produces a very strong and
fast blast wave that propagates outward and quickly decelerates. Once
the radius of the blast wave $r_b$ becomes much larger than the size
of the charge $r_0$ the flow can be considered self-similar of the
first kind and the only two parameters that completely determine its
properties are the initial density $\rho_0$ and the initial charge
energy $E_0$. Structure of the post-shock flow is characterized by the
very steep drop in density and pressure behind the blast wave front
(e.g., see Figs.~\ref{Sedov2DImage} and \ref{SedovCut}). However,
while density drops to essentially vacuum in the central region,
pressure asymptotes to a constant and typically fairly high
value. With good approximation it can be said that pressure remains
nearly constant in the inner 50\% of the blast wave radius and this
inner region is typically called the ``pressure plateau''. Since
density asymptotically approaches zero inside pressure plateau,
temperature asymptotically tends to infinity toward the blast wave
origin. Fig.~\ref{Sedov2DImage} shows the contour plot of the density
logarithm in the computational domain for the run Sedov.2D.b.256.100
at the end of the simulation, while Fig.~\ref{SedovCut} shows the
density distribution along the diagonal cut of the computational
domain for all runs in 2D and 3D discussed in this work. The full
analytic solution for the structure of the flow can be found in
\citep{Korobeinikov, ZeldovichRaizer} and we refer to those works for
further details.

As it was discussed in \S~\ref{GeneralFormalism}, problems involving
stationary ambient medium which is dynamically important, including
propagation of global shocks, are not optimal applications of the
method presented here. However, the extremely demanding conditions
presented by this problem for the numerical codes as well as the
availability of the analytic solution make this an excellent test
problem, which we use to verify the accuracy of our method as well as
its convergence properties in the case of flows with
discontinuities. The ability of the scheme to converge to the correct
analytic solution in this case is crucial to demonstrate the fact that
the non-conservative nature of the method does not introduce a
systematic error to the solution and the Rankine-Hugoniot conditions
are valid in the transformed reference frame $\mathbf{\tilde X}$. As
was discussed in \S~\ref{EqnInvariance}, such verification of validity
of shock jump conditions would be even more critical in the cases of
non-trivial choices of scaling parameters $\alpha$ and $\beta$ when
rigorous analytical proofs of such validity are not available.

This class of tests was carried out with all types of computational
reference frames discussed above, except for \emph{type c} (constant
acceleration contracting frame). All tests were compared against the
full analytic solution \citep{Korobeinikov} as well as against the
results of a corresponding reference simulation performed in the
inertial frame. We also carried out this test in three dimensions in
the inertial and constant acceleration (type b) frames in order to
verify the accuracy of the method in 3D. All 2D and 3D runs, including
the ones in the inertial frame, had the same total physical simulation
run time $t_{tot}$ (see Table~\ref{Runs2}). Moreover, it was ensured
that in all moving frame runs the domain extent in physical space at
the end of the simulation coincided with that of the simulation in the
inertial frame. To achieve that in the case of oscillating frame runs
the number of oscillation periods was always integer and, thus, the
domain extent in physical space was the same at the beginning and the
end of the run. This guaranteed that the effective resolution of each
run, i.e., cell size in physical space at the finest refinement level,
was the same at time $t=t_{tot}$, when cross-comparison of the runs
was performed.

Initial conditions in all runs (both 2D and 3D) are as follows.
Ambient density is $\rho_0 = 1.0$ and ambient pressure is
$P_0=10^{-4}$. It should be noted that in order to reproduce the true
Sedov blast wave the ambient pressure would have to be set at
essentially zero value. While this is possible in the case of a
stationary reference frame, in which ambient material velocity is
zero, in a moving reference frame that is impossible. The reason for
that is the large velocity $\vect{\tilde u} = -\dot{a}\vect{\tilde r}$
of the ambient material in the computational frame which is necessary
to support the shock. The fact that 100\% of the total energy in the
ambient material is kinetic energy results in breakdown of the
solution in the hydrodynamic solver in cells containing ambient
material. On the other hand, the maximum difference (max-norm) between
the lowest resolution runs performed in a stationary frame with
$P_0=10^{-4}$ and $P_0=10^{-16}$ is $1.5\cdot10^{-6}$.  This is almost
4 orders of magnitude less than the 1-norm error between the numerical
and exact solution (cf. Fig.~\ref{SedovConvergence}). Hence, we
conclude that at the resolutions considered the solutions obtained
with $P_0=10^{-4}$ are virtually identical to the solutions that would
be obtained with $P_0 = 0$. The charge is a cell at the finest
refinement level located in the corner of the domain with coordinates
$x_i = 0$. Charge density is $\rho_0$ and charge energy is $E_0 =
1000.0$. Initially fluid is at rest in the inertial frame.

All runs were performed in a quadrant in 2D and an octant in 3D. Due
to this boundary conditions on the lower $x$, $y$ (and $z$) boundaries
were reflective. Boundary conditions on the upper $x$, $y$ (and $z$)
boundaries where of the type appropriate for the problems with
dynamically important ambient material, as discussed in
\S~\ref{NumMethod}.

\emph{2. Converging shock (Guderley blast wave)}

We use the problem of a converging shock, i.e., the so-called Guderley
blast wave, as an example of a collapsing environment. In this problem
a strong spherical or cylindrical shock is initiated by some
mechanism, e.g., a piston or simply a pressure jump in the initial
conditions. The shock propagates toward the center of symmetry of the
system increasing its strength, i.e., undergoing cumulation, until the
moment of collapse. Thus, this problem presents the same complication
for the method discussed here as the strong point explosion since the
collapsing shock propagates in the medium stationary in the inertial
frame. The solution of this problem was first obtained by
\citep{Stanyukovich,Guderley} and we refer to \citep{ZeldovichRaizer}
for the detailed discussion (see also references therein). The
converging shock is an example of a self-similar problem of the second
type. In such problems the value of the similarity exponent $\kappa$
must be found based on the limiting self-similar solution that exists
close to the instant of collapse. There is no general analytic form of
such limiting solution, consequently it must be determined
numerically. In particular, as the shock wave radius decreases, the
solution in the region, the radius of which is of the order of the
shock radius and is proportional to it, will be approaching the
limiting solution thereby giving an approximation of the latter. The
only dependence of that limiting solution on the initial conditions
will be described by the parameter $A$, which characterizes the
intensity of the initial push. The second unique property of a
self-similar problem of the second type is the existence of a critical
characteristic in the $r,t$ plane, which is of the same family as the
shock wave characteristic and which converges with the latter at the
moment of collapse. That characteristic defines the region of
influence, i.e., the shock wave cannot be affected in any way by the
flow outside the region bounded by the critical characteristic and,
thus, it does not depend on the outer boundary conditions. Therefore,
in solving the problem of the converging shock wave, it is very
important to obtain the structure of the flow in the vicinity of the
shock front as accurately as possible and as closely to the instant of
collapse as possible as that structure is then used to represent the
sought limiting solution based on which all of the characteristics of
the flow are obtained. Extremely demanding conditions presented by
this test for a numerical code, in particular the sharp rise in
pressure and temperature and, consequently, the shock strength near
the moment of collapse, as well as the availability of a single
parameter that is highly sensitive to the quality of the solution,
namely the similarity exponent $\kappa$, make this an excellent test
for assessment of the performance of this method in the case of
contracting reference frames.

\begin{figure}[t]
\epsscale{1.05}
\plottwo{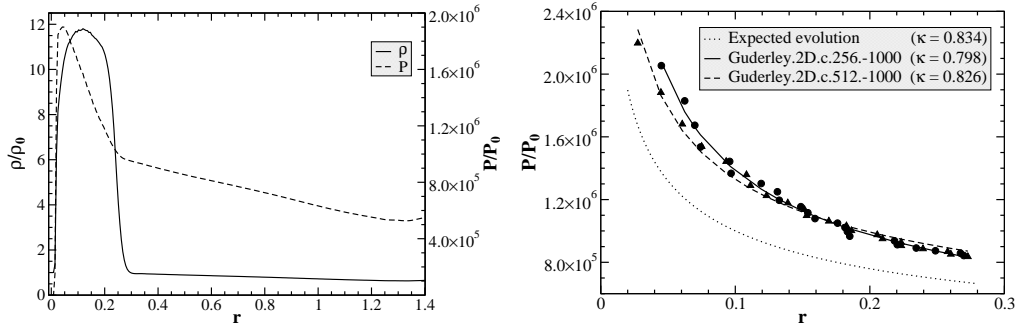}{fig3b.eps}
\caption{\emph{Left:} Structure of the converging shock wave right
before the shock front collapse at the center of symmetry. Shown is
density and pressure, normalized to their pre-shock values $\rho_0$
and $P_0$, at time $t = 0.89\cdot 10^{-3}$. Profiles are taken along
the diagonal cut of the computational domain of the run
Guderley.2D.S.256. \emph{Right:} Position and values of pressure
maxima for the final stages of collapse of the converging
shock. Circles represent pressure maxima at each moment in time for
the run Guderley.2D.c.256.-1000, while the triangles show the same for
the run Guderley.2D.c.512.-1000. The solid and dashed lines show the
corresponding fits for those two runs. The dotted line shows the
expected evolution.
\label{Guderley2D}}
\end{figure}

All performed simulations of this type of test were carried out in a
constant acceleration contracting frame, i.e., \emph{type c} as
discussed in \S~\ref{FrameTypes}. As with other tests, we also
performed a reference run in the inertial frame (see
Fig.~\ref{Guderley2D}). The total duration in physical time of all
simulations as well as their initial domain extent in physical space
is the same (see Table~\ref{Runs1}). At the end of the moving frame
runs their domain extent in physical space is exactly half that of the
reference run, therefore, the final resolution of the run
Guderley.2D.c.256.-1000 is twice the resolution of the reference run
Guderley.2D.S.256.

Initial distribution of all physical quantities, including fluid
velocities, is identical in all runs. Initial density is uniform
$\rho_0 = 1.0$. Pressure distribution contains a jump along the
interface with the radius $r_P = 0.8$. Pressure inside the interface
is $P_0 = 1.0$, while outside is $P_{out} = 10^6$. Here, as in the
case of the Sedov blast wave, the pre-shock pressure should ideally be
set to zero. However, for the same reason as the one discussed before
that is impossible in the case of a contracting reference frame.
Moreover, in this case we also find that the significant decrease in
pre-shock pressure has a negligible effect on the solution. All runs
were performed in a quadrant in 2D. Consequently, boundary conditions
on the lower $x$ and $y$ boundaries were reflective. As it was said
above, the choice of boundary conditions on the upper $x$ and $y$
boundaries does not affect the structure of the flow in the shock
vicinity. Thus, we use zero-order extrapolation boundary conditions on
the upper $x$ and $y$ boundaries, as it was discussed in
\S~\ref{NumMethod}.

Left panel of Fig.~\ref{Guderley2D} shows the structure of the flow
right before the collapse of the shock front at the center of
symmetry. The region, representing the limiting self-similar solution,
extends to $r \approx 0.1$. The flow outside that region is completely
determined by the initial and boundary conditions. As the shock
approaches the point of collapse the pressure and temperature behind
the front tend to infinity, while density behind the front stays
constant and equal to $[(\gamma + 1)/(\gamma -1)]\rho_0$. Further
behind the front density rises monotonically with radius and
eventually it asymptotes to a limiting value $\rho_{lim}$, which is
achieved at the moment of collapse. For $\gamma = \sfrac{7}{5}$ the
limiting density is $\rho_{lim} = 21.6\rho_0$.

The similarity exponent describes the shape of the distribution of
basic quantities in the self-similar region immediately behind the
shock front. For example, the pressure behind the front is
\beq
P \sim R^{2(\kappa - 1)/\kappa},
\label{Pdependence}
\eeq
where $R$ is the current shock position. Monotonicity of pressure
behind the front depends on the value of the polytropic index
$\gamma$. For $\gamma = \sfrac{7}{5}$, used in our simulations, the
pressure rises behind the front until it reaches the maximum, after
which it monotonically decreases. Thus, in practice in a numerical
solution it is much easier to identify the pressure maximum, rather
than the value of pressure immediately behind the shock front.
Moreover, as can be seen in Fig.~\ref{Guderley2D}, such pressure
maximum trails the shock front rather closely. Right panel of
Fig.~\ref{Guderley2D} shows positions and values of pressure maxima
for a sample of times in the runs Guderley.2D.c.256.-1000 and
Guderley.2D.c.512.-1000. The moments in time were chosen fairly close
to the point of collapse when it is possible to assume that the
solution has a large degree of self-similarity. Subsequently, a fit
was produced for each run according to eq. (\ref{Pdependence}), based
on which the similarity parameter $\kappa$ was determined. The
obtained values of $\kappa$ for both runs are shown in the legend.
The dotted line shows the expected evolution of maximum pressure, and
in the legend the exact theoretical value of $\kappa$ is
given\footnote{Note that the expected evolution curve is shifted down
for clarity and is intended only to indicate the shape of the curve
rather than its absolute values.}.

\begin{figure}
\epsscale{1.0}
\plotone{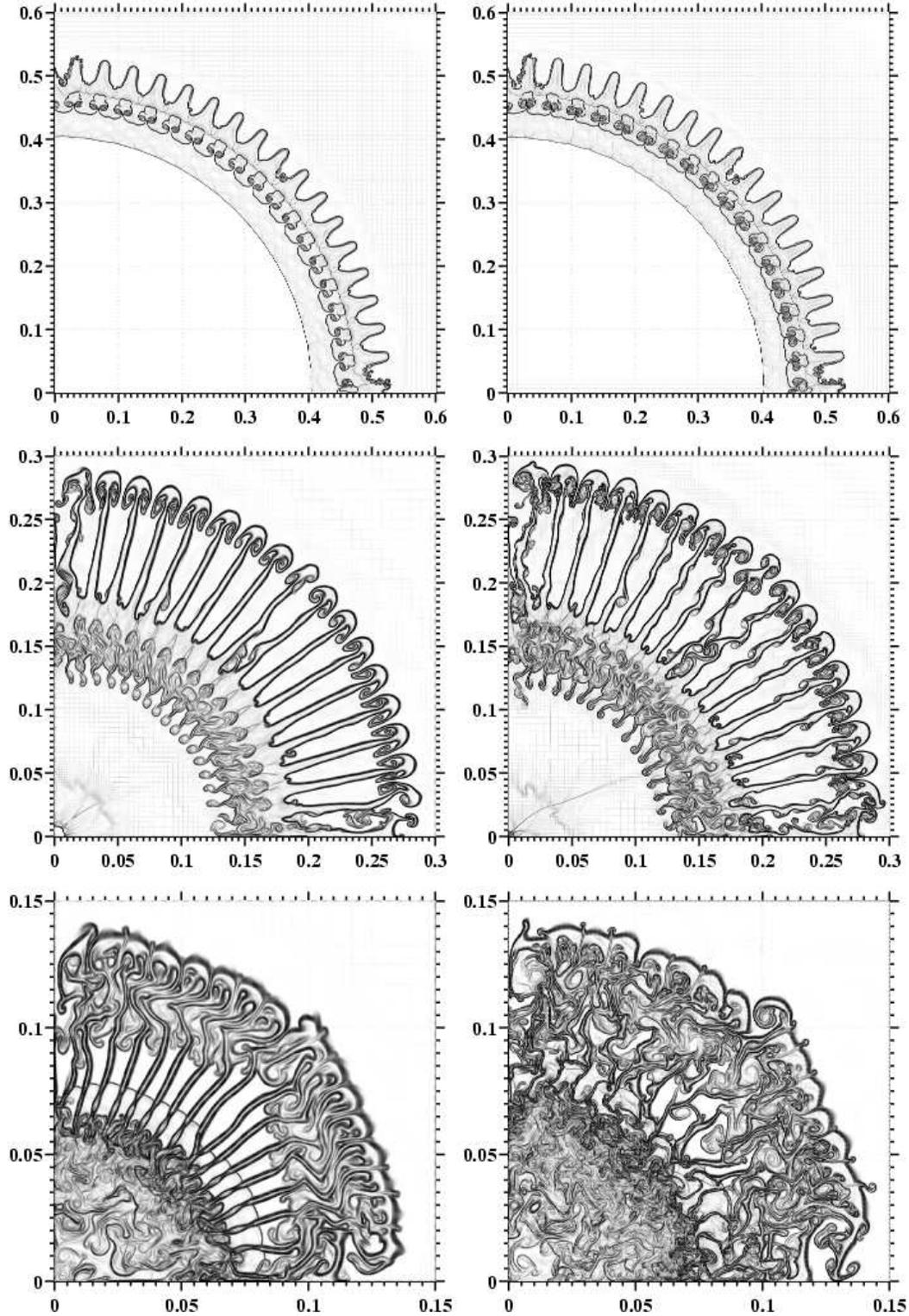}
\caption{Evolution of the converging shock wave interacting with a density
interface with imprinted perturbations. Shown are the Schlieren images
of the computational domain of the simulations GuderleyShell.2D.S.4096
(left panels) and GuderleyShell.2D.c.4096.-1250 (right panels) for
three moments in time, namely from top to bottom: $t = 0.358\cdot
10^{-3}$, $t = 0.576\cdot 10^{-3}$, and $t = 0.713\cdot 10^{-3}$.
\label{GuderleyPert}}
\end{figure}

We also conducted a more complicated variation of this test,
designated as ``GuderleyShell'' in Table~\ref{Runs1}. In that test a
converging shock, identical to the one discussed above, interacts with
the density interface with imprinted perturbations. This problem is
somewhat analogous to the collapse of a fuel pellet in the inertial
confinement fusion applications, in which the perturbations are the
result of surface nonuniformities of the pellet itself as well as of
the nonuniformity of the pellet surface illumination with laser
beams. The setup of this test is identical to the one described above,
except that besides the pressure jump at $r_P = 0.8$ there is also a
density jump at $r_{\rho} = 0.7$ with density inside that interface
$\rho_{in} = 0.05$. The interface itself is imprinted with a
sinusoidal perturbation with amplitude $A = 0.015 \approx 2\%
\, r_{\rho}$ and 18 perturbation periods in a quadrant. We performed 
a reference run in the inertial frame and a run in the constant
acceleration contracting frame. Computational domain setup and domain
resolution were identical in both runs at time $t = 0$. 

The key feature of this problem is the rapid inward growth of
perturbations driven by the Rayleigh-Taylor instability due to the
presence of dense pressurized shocked material above the interface.
Fig.~\ref{GuderleyPert} \citep{http} shows the initial stage of the
perturbation growth right after the shock impact of the density
interface, the intermediate stage, when the perturbation spikes are
fully evolved, and the final stage right after the moment of shock
front collapse. We use this test to illustrate the difference in the
final state of the system due to the two-fold increase in resolution,
which is provided by the reference frame contraction in otherwise
identical simulations.  Such resolution increase is primarily
manifested in the flow being much more unstable in the accelerating
frame run. Moreover, higher resolution of the central region allowed
to achieve higher central peak pressure at the moment of the shock
front collapse and, as a result, a faster rebound blast wave
propagating outward through the material continuing to collapse. It
should be noted that, as it was discussed in
\S~\ref{GeneralFormalism}, in this type of problems involving
propagation of a global shock the increase of the physical time step
cannot be expected since while the shock velocity is minimal in the
contracting reference frame, pre-shock material has large velocity in
the computational frame.  Nevertheless, in the case of the moving
frame with only minimal additional computational effort, primarily due
to the $\approx 10\%$ increase in the number of time steps, it was
possible to achieve the result that would require twice higher
resolution.

\emph{3. Expansion of a gas sphere into vacuum}

This problem, which is an example of the optimal application of the
method presented here, was used to demonstrate efficiency and long
term performance of the method in systems that exhibit large degree of
expansion. We also use these tests to verify the conservativity
properties of the method in a setting fairly representative of its
typical realistic application. We consider both an initially
stationary (designated as test category ``Sphere'' in
Table~\ref{Runs1}) and an initially rotating sphere (test category
``SphereRot'' in Table~\ref{Runs1}). The latter case allows us to
demonstrate the performance of the method with an expanding and
rotating reference frame. In all those tests the initial setup of the
problem is a sphere of radius $r_s = 0.3$ with constant initial
density $\rho_s = 10^3$ and constant initial pressure $P_s =
10^5$. Ideally ambient conditions for this problem should be
vacuous. However, since that is unfeasible in a Eulerian code, ambient
conditions were set to have the minimal possible dynamical effect on
the expansion of a sphere. In particular, in the reference runs
performed in the inertial frame ambient density was $\rho_0 = 10^{-2}$
and ambient pressure was $P_0 = 10^{-2}$ in the non-rotating case and
$\rho_0 = 10^{-3}$ and $P_0 = 10^{-3}$ in the rotating case. In the
moving frame runs ambient material was set to be stationary in the
computational frame and, therefore, it was expanding along with the
sphere material. Consequently, it was possible to use higher values
for the ambient density and pressure without causing dynamical effects
on the expansion of the sphere, namely $\rho_0 = 1.0$ and $P_0 = 1.0$
in the non-rotating case and $\rho_0 = 0.1$ and $P_0 = 0.1$ in the
rotating case.

\begin{figure}[t]
\epsscale{1.07}
\plottwo{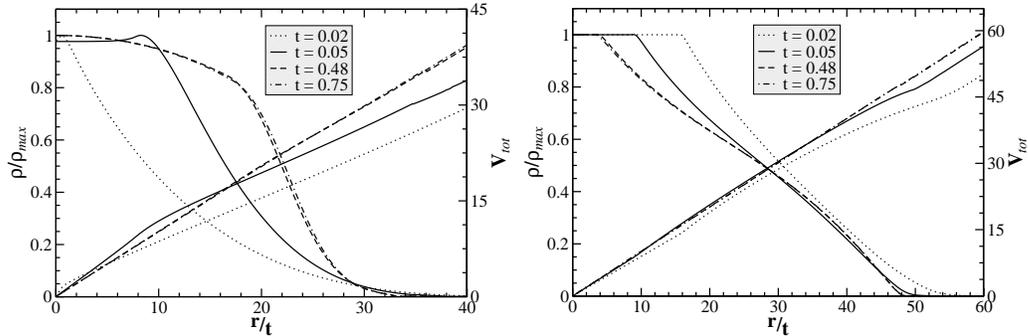}{fig5b.eps}
\caption{\emph{Left:} Run Sphere.2D.a.2048.50. \emph{Right:}
Run SphereRot.2D.$\textrm{e}_2$.2048.80. Both panels show normalized
density (set of decreasing curves) and total velocity (set of
increasing curves). Density is normalized to its maximum value at each
time. All distributions are given along the diagonal cut of the
domain.
\label{Sphere2D}}
\end{figure}

\begin{figure}[t]
\epsscale{0.53}
\plotone{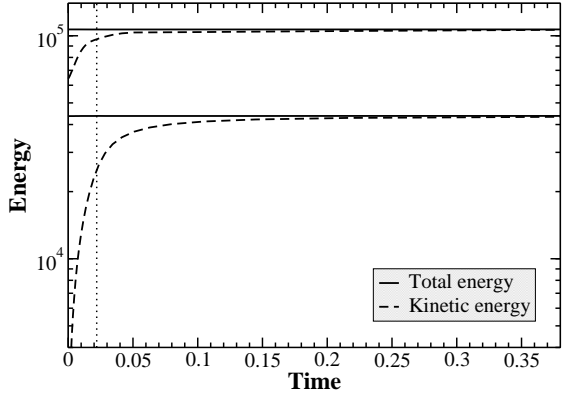}
\caption{Temporal evolution of the kinetic energy in the case of expansion
of a non-rotating sphere (lower pair of curves, run
Sphere.2D.a.2048.50) and a rotating sphere (upper pair of curves, run
SphereRot.2D.$\textrm{e}_2$.2048.80) into vacuum. Shown only the first
half of the total physical simulation run time. Vertical dashed line
corresponds to the total run time of the runs Sphere.2D.S and
SphereRot.2D.S.
\label{SphereEnergyEvolution}}
\end{figure}

Fig.~\ref{Sphere2D} shows the solution structure for the non-rotating
and rotating sphere cases at four different times\footnote{Note, that
both panels exclude the outer part of the computational domain that
does not contain sphere material.}. Both density and velocity are
shown as functions of the similarity variable $r/t$ thereby
illustrating gradual development of the self-similar structure by the
flow.  Fig.~\ref{SphereEnergyEvolution} shows the temporal evolution
in the system of the total kinetic energy measured in the inertial
frame for two cases.

In the non-rotating case expansion starts with the outer layer which
expands with the maximum escape velocity $u_{max} = [2/(\gamma -
1)]c_s$, where $c_s$ is the initial sound speed in the sphere. For the
conditions given above $u_{max} = 38.7$. The rarefaction wave
propagates toward the sphere center. The time when it reaches the
center marks the end of the acceleration phase. At that point all of
the sphere material has been disturbed and the linear velocity profile
sets in. The dotted lines in Fig.~\ref{Sphere2D} show the density and
velocity profiles at that time, i.e., at $t = 0.02$. Note that, as can
be seen in Fig.~\ref{SphereEnergyEvolution}, at the end of the
acceleration phase in the non-rotating case kinetic energy constitutes
about 56\% of the total energy. After that the rarefaction wave
bounces off at the center and the reflected rarefaction starts to
propagate outward. This phase is shown with the solid lines in
Fig.~\ref{Sphere2D} for $t=0.05$. Initially the flow in the system is
not self-similar since there is a characteristic dimension in the
problem, namely the initial sphere radius $r_s$. However, as the
expansion proceeds and the flow extent becomes much larger than $r_s$,
it eventually ``forgets'' about the initial conditions and the flow
asymptotically approaches the self-similar regime. This can be seen in
Fig.~\ref{Sphere2D} as the density and velocity distributions at
$t=0.48$ and $t=0.75$ are almost identical.

In the rotating case the initial angular velocity of the sphere is
$\omega_s = 100.0$ (see Table~\ref{Runs1}). This value was chosen to
ensure that initially kinetic and thermal energies of the sphere
material are comparable, with the initial kinetic energy due to
rotation constituting about 60\% of the total energy (see
Fig.~\ref{SphereEnergyEvolution}). Evolution of the rotating sphere
critically depends on the magnitude of that initial kinetic energy.
Our choice of the angular velocity ensured, on one hand, an important
role of the acceleration phase during which the rarefaction wave
traveled into the sphere interior and the kinetic energy rose from
60\% to about 95\% of the total energy (see
Fig.~\ref{SphereEnergyEvolution}). The dotted lines in the right panel
of Fig.~\ref{Sphere2D} show the flow structure at the end of that
phase, i.e., at $t=0.02$. On the other hand, the rarefaction wave
never reached the center and, thus, the reflected rarefaction was
never produced as can be seen in the absence of the characteristic
bump in the solid line showing density at $t = 0.05$ (cf. the solid
line for the same time in Fig.~\ref{Sphere2D}). Note also that in the
rotating sphere the expansion velocity is higher than in the
non-rotating one due to the centrifugal force acting on the rotating
fluid. Consequently, the flow tends to ``forget'' the initial
conditions much sooner and, therefore, to approach the self-similar
regime much more rapidly.

In performing those tests we used only the reference frames that are
most suited for this type of problems and can provide the optimal
performance in the absence of fluid motions feedback. For both the
non-rotating and rotating cases we also carried out a series of
reference runs in the inertial laboratory frame. In the non-rotating
case the most natural choice for the reference frame was the constant
velocity expanding frame (runs Sphere.2D.a in Table~\ref{Runs1}).
During the acceleration phase, however, the linear velocity profile
gradually extends inward along with the rarefaction wave propagating
inward. In the computational frame that results in the sphere material
initially possessing a large velocity gradient $\vect{\tilde u} =
-\dot{a}\vect{\tilde r}$ (see eq. (\ref{newvel})) which is slowly
being eliminated as the fluid accelerates until all of the sphere
material is comoving with the reference frame. The initial presence of
such velocity gradient can cause the degradation of the solution. In a
realistic application a feedback mechanism would cause the reference
frame to accelerate gradually until the linear velocity profile is
established thereby eliminating such a problem. In order to assess the
improvement of the solution due to such more accurate tracking of
fluid motions, we carried out a set of simulations utilizing the
delayed stretch reference frame, i.e., type $\textrm{a}_d$ as
discussed in \S~\ref{FrameTypes}, designed to imitate the action of
such a realistic feedback mechanism (runs Sphere.2D.$\textrm{a}_d$ in
Table~\ref{Runs1}). Those simulations were initialized by transforming
the computational domain of the reference inertial frame runs at time
$t_s = 0.021$ to the constant velocity expanding frame.

In the case of a rotating sphere angular velocity of the fluid
resulted in all of it having a radial expansion velocity from $t =
0$. Therefore, with a good approximation throughout the acceleration
phase the interior of the sphere could be characterized by the linear
velocity profile\footnote{Note that the break in the velocity profile
around $r/t = 50$ in Fig.~\ref{Sphere2D} at the two earliest times is
the reflection of the presence of the mismatch still existing at that
time between the expansion velocity of the sphere material and the
grid velocity.}. That justified the use of an expanding frame for the
whole duration of the runs and eliminated the need for the use of the
delayed stretch. At the same time in the process of expansion the
fluid quickly loses its angular velocity due to the conservation of
angular momentum. In order to account for that we used the constant
velocity expanding frame with expansion-correlated angular velocity,
i.e., type $\textrm{e}_2$ as discussed in \S~\ref{FrameTypes} (runs
Sphere.2D.$\textrm{e}_2$ in Table~\ref{Runs1}). In both the
non-rotating and rotating cases the grid velocity $v_g$ was chosen to
be slightly lower than the expansion velocity of the sphere. For
example, in the runs Sphere.2D.a velocity $v_g$ of the grid right
boundaries with coordinates $\tilde r_b = 0.6$ (see Table~\ref{Runs2})
was 50.0 (see Table~\ref{Runs1}). Thus, initially at the sphere
boundary $\tilde r_s = 0.3$ grid velocity was $v_{g,s} = (v_g/\tilde
r_b)\tilde r_s = 25.0$ which is lower than the maximum expansion
velocity into vacuum $u_{max} = 38.7$. Therefore, sphere material in
all moving frame simulations initially expanded somewhat until the
outer layer of the sphere reached the point in the domain with
matching grid velocity.

The non-rotating sphere runs were performed in a quadrant with the
sphere center located in the lower left corner of the domain. In the
rotating case, the domain contained the full sphere in order to
accommodate its rotation with the sphere center and the fixed point of
reference frame expansion located in the center of the computational
domain. We carried out each type of runs with a range of resolutions.
The total time $t_{tot}$, given in Table~\ref{Runs2}, of the reference
runs performed in the inertial frame was chosen to cover the duration
of the acceleration phase of the expansion. Correspondingly, the
domain size in those runs was set to accommodate the expansion of the
sphere during that time. In the moving frame runs, with the exception
of the delayed stretch runs Sphere.2D.$\textrm{a}_d$, the domain size
was chosen to be half the size of the domain in the reference runs,
since the sphere does not expand significantly in the computational
frame. Therefore, at $t = 0$ the computational domain in the moving
frame coincides with the inner half of the domain of the corresponding
reference runs. Thus, initially cell size in physical and
computational space is identical for all runs at a given resolution
level.

Finally, since ambient material in all simulations was set to be
stationary in the current reference frame, boundary conditions in the
case of a non-rotating sphere were of the reflective type on all
domain boundaries as discussed in \S~\ref{NumMethod}. This allowed us
to use those runs for the study of the conservativity properties of
the method in \S~\ref{Conservativity} below. In the simulations of a
rotating sphere boundary conditions were of the zero-order
extrapolation type on all boundaries.

\emph{4. Isentropic expansion with an embedded density structure}

The fourth type of tests we conducted is the isentropic expansion of a
uniform pressure field with an embedded density structure (test
category ``Clump'' in Table~\ref{Runs1}). The initial setup of the
problem is the computational domain with constant initial pressure
$P_0 = 1.0$. The embedded density structure is a circular clump with
uniform density $\rho_c = 10^3$ while ambient density is $\rho_0 =
1.0$.  Clump radius is $r_c = 0.15$ and its center has coordinates
$\tilde x_i = 0.3$. All fluid has constant expansion velocity $ u_i =
\dot{a}\tilde x_i$, where $\dot{a}$ is such that both components of
the clump center velocity are equal to $50.0$. The computations are
performed in the reference frame co-expanding with the fluid with the
center of expansion located in the lower left corner of the domain.
Consequently, all boundaries have perfectly reflecting boundary
conditions.

In the course of its evolution the domain expands by more than two
orders of magnitude while pressure drops to $10^{-7}$ and clump
density drops to $0.063$. In this problem the exact structure of the
flow at time $t_{tot}$ can be determined analytically, moreover the
clump boundary must remain as a sharp discontinuity. Consequently, we
use this test to verify the second order convergence of the numerical
solution to the exact one.

\subsection{Test results: method accuracy}
%-------------------------------------------------------------------------------
\label{Accuracy}

There are two principal sources of errors in the strategy employed by
us for solving the system (\ref{EulerTr1}) - (\ref{EulerTr3}), i.e.,
operator splitting. They are, on one hand, the numerical errors
introduced individually by the hydrodynamic and source term solvers
and, on the other hand, the errors due to the imbalance that can arise
during a time step between the two solvers.

In order to illustrate the origin of the first source of errors
consider the action of the two operators in the course of one time
step in the case of direct operator splitting as described by the
eq. (\ref{OperatorSplitting}). It is convenient to rewrite that
equation as follows
\beq
\begin{array}{rcl}
\tilde q^{n+1} = \mathcal{S}^{(\Delta t)}\mathcal{H}^{(\Delta t)}\tilde q^n & \, = \, &
\mathcal{S}^{(\Delta t)}\left( \tilde q^{*,n+1}_H + \delta \tilde q^{n+1}_H\right) = \nonumber \\
& & \tilde q^{*,n+1} + \delta \tilde q^{n+1}_S + \delta \tilde q^{n+1}_{HS}
+ \delta^2 \tilde q^{n+1}_{HS}.
\end{array}
\label{SolutionError}
\eeq
The solution is first advanced over the full step $\Delta t$ with a
hydrodynamic solver. The resulting solution consists of the exact
solution $\tilde q^{*,n+1}_H$ of the homogeneous part of the original
equations and the numerical error $\delta \tilde q^{n+1}_H$ introduced
by the solver. That error does not violate the conservativity of the
solution, provided that the hydrodynamic scheme is conservative. After
that source terms are applied adjusting the obtained solution to
account for the effects of grid expansion/contraction and to apply
forces due to the grid acceleration and/or rotation. Thus the final
solution at the end of the time step consists of the exact solution
$\tilde q^{*,n+1}$ of the full equations and the error $\delta \tilde
q^{n+1}_S$ produced by the source term solver acting on the exact part
$\tilde q^{*,n+1}_H$ of the homogeneous solution, as well as the
errors $\delta \tilde q^{n+1}_{HS}$ and $\delta^2 \tilde q^{n+1}_{HS}$
due to the source term solver action on the error $\delta \tilde
q^{n+1}_H$. The error $\delta \tilde q^{n+1}_{HS}$ is the ``exact''
part of the result of the application of the source term operator
$\mathcal{S}$ to the hydrodynamic error $\delta \tilde q^{n+1}_H$,
therefore, it does not violate the conservativity of the final
solution. That is not the case for the other two errors since the
source term solver simply adds or subtracts a certain amount of state
vector components in each cell. A small numerical error in that
amount, i.e., the sum of errors $\delta \tilde q^{n+1}_S$ and
$\delta^2 \tilde q^{n+1}_{HS}$, results in violation of
conservativity. As was discussed in \S~\ref{NumMethod}, we use the
second-order accurate conservative explicit hydrodynamic schemes and
the fourth-order accurate implicit source term solver. Consequently,
the error $\delta \tilde q^{n+1}_{HS}$ has the largest effect on the
accuracy of the solution. It drops quadratically with the decrease in
spatial and temporal step. The error $\delta \tilde q^{n+1}_S$ is
significantly smaller due to the much higher accuracy of the source
term solver, moreover it decreases as the $4^{th}$ power of the
spatial and temporal step. Finally, the error $\delta^2 \tilde
q^{n+1}_{HS}$ has the smallest magnitude. It does not have a unique
dependence on the spatial and temporal steps since the ``exact''
solution $\delta \tilde q^{n+1}_{HS}$ itself changes with resolution.

Aside from employing more accurate hydrodynamic schemes or using the
brute force approach of higher resolution, the most efficient way to
minimize all of the above errors is by closely correlating the motion
of the non-inertial frame with the fluid motion and, thereby, by
minimizing the velocity field in the computational domain. However,
that will have the largest effect on the error $\delta \tilde
q^{n+1}_{HS}$ since it typically dominates other errors. In the limit
of the reference frame perfectly following the fluid motion so that
the material is stationary in the computational domain, the
hydrodynamic solver does not alter the solution during a time step, as
no waves are produced at cell interfaces, and, therefore, it does not
produce the error $\delta \tilde q^{n+1}_H$. Consequently, the error
$\delta^2 \tilde q^{n+1}_{HS}$ is also absent. However, the error
$\delta \tilde q^{n+1}_S$ would still be present. In
\S~\ref{Conservativity} and \S~\ref{Convergence} we illustrate the
effect this can have on the conservativity and convergence properties
of the solution. The first and last errors in eq.
(\ref{SolutionError}) can be further minimized by increasing the
accuracy of the source term solver. One of the major advantages of the
Kaps-Rentrop method, used by us, is the fact that this scheme is
embedded, i.e., it provides means to control the solver accuracy or,
equivalently, the magnitude of the relative error of the solution. In
all simulations discussed in this work we use the target value of the
relative error of $10^{-4}$. We found it to be the most optimal
compromise between speed and accuracy of the solver. However, even
infinite accuracy of the latter would not completely eliminate the
error $\delta\tilde q_{HS}^{n+1}$ since there is always a seed present
$\delta \tilde q^{n+1}_H$.

The second source of numerical errors is the imbalance between the
action of the hydrodynamic and source term operators during a time
step. It can be seen from the eqs. (\ref{EulerTr1}) - (\ref{EulerTr3})
as well as the expressions for $d\ln a/d\tau'$, $d^2 \ln a/d\tau'^2$,
and $d\ln\Omega/d\tau'$ given in \S~\ref{FrameTypes} that source terms
usually depend explicitly on time, i.e., grid velocity $\dot{a}(t)$
can vary substantially in the course of one time step. We found that
two techniques can be very efficient in maintaining balance between
the two operators. Firstly, in \S~\ref{NumMethod} it was discussed
that in our source term solver we use time step subcycling.
Consequently, in the course of such subcycling source terms and their
Jacobian $A_{ij}$ at each substep must be evaluated using the value of
$t$ that corresponds to the beginning of the global time step. This
also applies to both substeps in the case of Strang splitting. The
second technique is limiting the time step, at which the simulation is
advanced, in correspondence with the rate of change of the source
terms, namely
\beq
d\tau = \displaystyle \alpha \min\left\{d\tau_H,\left |
\tilde q_p  \left(\frac{d\tilde q_p}{d\tau'}\right)^{-1} \right |\right\},
\label{dtLimited}
\eeq
where $d\tau_H$ is the next time step determined based on the CFL
condition (\ref{CFLmovingframe}) and the factor $\alpha$ is typically
$0.05 - 0.1$. Subscript $p$ indicates the state vector components that
represent density and energy. We do not use momenta in the
determination of the time step as they can have zero values. Values
for $d\tilde q_p/d\tau'$ are simply found based on the right-hand side
of the original equations. The above prescription ensures that source
terms do not vary significantly within one time step thereby allowing
the hydrodynamics to adjust appropriately to the changes in the
velocity of the reference frame.

AMR can serve as an additional source of errors when an insufficiently
accurate interpolation technique is used in the process of refining a
certain region. Consider the fluid of uniform density and pressure
stationary in the inertial reference frame. Assume that the
computational domain resolution is increased from $dx$ to
$dx/2$. While density and pressure are uniform and constant throughout
the domain, the velocity gradient $\vect{\tilde u} =
-\dot{a}\vect{\tilde r}$ would have to be interpolated appropriately
to the new resolution. Assume that constant interpolation is
used. Then in two adjacent fine cells, which replaced one coarse cell,
velocity will be the same instead of representing the velocity
gradient. This will lead to the creation of artifact features in the
computational domain. The use of higher order interpolation
techniques, e.g., centered linear interpolation, significantly removes
such artifacts although at a certain level they are always present.

\begin{figure}[t]
\epsscale{1.07}
\plottwo{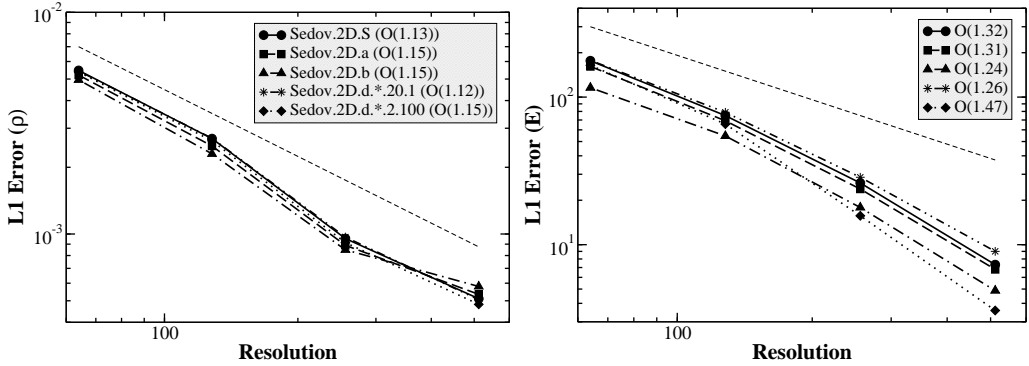}{fig7b.eps}
\caption{Convergence of the numerical solution to the exact one in simulations
of the Sedov blast wave in various reference frames. \emph{Left:} $L1$
error in density. \emph{Right:} $L1$ error in total energy in the
laboratory reference frame. The order of convergence is shown in
brackets in the legend in both panels. The dashed lines correspond to
first-order convergence.
\label{SedovConvergence}}
\end{figure}

Figs.~\ref{Sedov2DImage}, \ref{SedovCut} and \ref{SedovConvergence}
illustrate the overall accuracy of the method presented here for the
case of the strong point explosion test, discussed in
\S~\ref{TestTypes}. Note in Fig.~\ref{Sedov2DImage} that the overall
sphericity and azimuthal uniformity of the solution is preserved and
no artifacts along the axes were introduced neither by the
hydrodynamic scheme, nor due to the fact that the simulation was
performed in the expanding non-inertial reference frame. The only
exception is the central vacuous region in which internal energy
becomes extremely high. However due to the presence of singularity at
the center in the exact solution the numerical solution in that region
obtained with practically any Eulerian code would significantly differ
from the analytic one. Fig.~\ref{SedovConvergence} shows the
convergence of the $L1$ error in density and total energy between the
numerical and exact solutions in simulations of the Sedov blast
wave. It can be seen in Fig.~\ref{SedovConvergence} (see also
Fig.~\ref{SedovCut}) that even for a fairly low resolution of 64 cells
the solution accuracy in the moving frame runs is rather
high\footnote{Note, that the errors shown are absolute and not
normalized.}. More importantly, it can be seen in
Fig.~\ref{SedovConvergence} that even in this test, which is not an
optimal application of this method, the solution accuracy practically
for all types of moving frames is equal to or greater than the
accuracy of the solution in the laboratory frame. The only exception
are the higher resolution runs performed in the high amplitude low
frequency oscillating reference frame in which the solution has a
slightly higher error in energy. That shows that the computation in a
moving frame does not introduce a systematic error and the numerical
solution properly converges to the exact one.

One important point, that follows from Figs.~\ref{SedovCut} and
\ref{SedovConvergence}, concerns the behavior of the runs performed in
the oscillating reference frames. It is unlikely that in realistic
applications the frame motion would be smooth as it is in the runs
with the constant velocity or constant acceleration frames.
Consequently, it is important to assess what level of oscillatory
noise is acceptable in the frame motion in order to determine how one
should set up the filtering in the fluid motion feedback
mechanism. The run Sedov.2D.d.256.20.1 was designed to simulate the
low frequency high amplitude oscillations of the frame while the run
Sedov.2D.d.256.2.100 was intended to simulate the high frequency low
amplitude noise. In both runs, in particular in the latter one, source
terms have a very high degree of temporal variability. Therefore, it
is essential to use time step limiting as described by
eq. (\ref{dtLimited}), without which the solution quality dramatically
degrades. With the proper use of such step limiting it is possible to
have very high frequency oscillations of the reference frame and still
have the solution to virtually coincide with the reference run
performed in the inertial frame. It is very important to note, though,
that the time step determined according to eq. (\ref{dtLimited}) can
be much lower than the maximum time step allowed for this method,
based on eq. (\ref{CFLmovingframe}), and that can adversely impact the
overall performance of the method. In fact, the run
Sedov.2D.d.256.2.100 took 110877 steps vs. 6365 steps for the run
Sedov.2D.d.256.20.1, 6512 steps for the run Sedov.2D.a.256.20 and 7756
steps for the run Sedov.2D.b.256.100. Therefore, while the method
discussed in this work allows one to accommodate high frequency
oscillations of the frame without any significant loss of accuracy, it
is beneficial for the code performance to filter out such high
frequency noise and only to follow smooth global motions of the
fluid. Of course, a healthy balance must be found between that and
still closely tracking the fluid motions since, as was discussed
above, the loss of such close correlation increases the error
introduced by the hydrodynamic solver.

Solution accuracy in the case of the contracting reference frame,
considered for the problem of the converging shock, is illustrated in
the right panel of Fig.~\ref{Guderley2D}. It can be seen that for the
lower resolution run the error in the similarity exponent $\kappa$,
determined based on the numerical solution, is about 4.3\%, while for
the run with twice higher resolution the error drops by more than a
factor of 4 and is slightly less than 1\%. Therefore, in such a
numerically challenging test, which is not an optimal application of
the method presented here, as was discussed before, it is possible to
achieve exceptional solution accuracy with contracting reference
frames even in the cases of modest resolution.

The above discussion of the method accuracy is also valid when solving
the system (\ref{EulerTr1}) - (\ref{EulerTr3}) with a different choice
of scaling parameters $\alpha$ and $\beta$, in particular the one
given by (\ref{BestChoice}) and providing conservation of mass. In
performing the same strong point explosion tests as the ones discussed
above with such choice of $\alpha$ and $\beta$ the accuracy is
marginally better due to the fact that the source term solver does not
introduce the error in density. In cases in which system evolution is
followed over much longer time periods such improvement in accuracy
can be more prominent. That effect can be even more pronounced for
less accurate source term solvers, in particular the ones with
accuracy comparable to that of the hydrodynamic solver. In those cases
the choice (\ref{BestChoice}) of $\alpha$ and $\beta$ ensuring
conservation of mass (and momentum in constant velocity frames) may
significantly improve solution accuracy. Scaling parameter choice can
also have an impact on the method performance due to the time step
restriction given by eq. (\ref{dtLimited}).  Indeed, the absolute
magnitude of the source terms and, thus, the rate of change of the
state vector may vary depending on the form of the source terms given
by different choices of scaling parameters. Finally, solution accuracy
can be further increased by the use of Strang splitting. However, in
tests discussed here we find this to lead to only marginal solution
improvement, moreover in discontinuous flows the convergence rate
remains first order in the case of Strang splitting, which, on the
other hand, can have significantly larger computational cost.

\subsection{Test results: conservativity properties}
%-------------------------------------------------------------------------------
\label{Conservativity}

\begin{figure}
\epsscale{1.09}
\plottwo{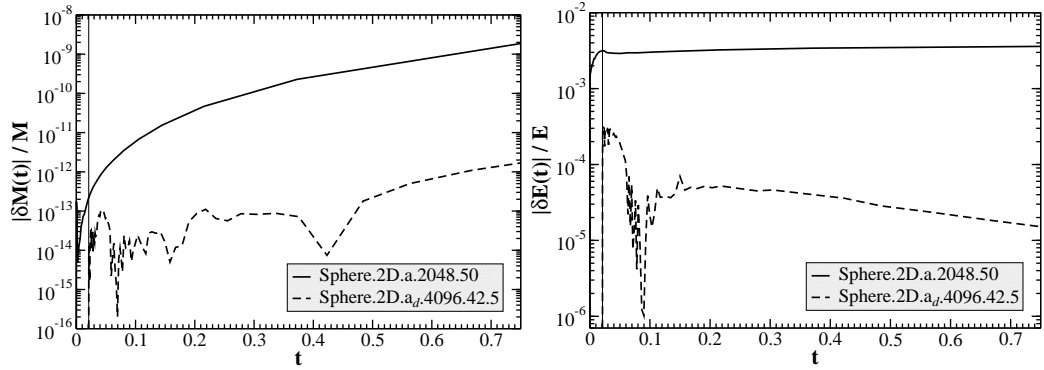}{fig8b.eps}
\caption{Temporal evolution of the relative error in total mass
(\emph{left}) and total energy (\emph{right}) conservation for the
simulations of expansion of a non-rotating sphere into vacuum.
Vertical line indicates the time of start of the reference frame
expansion for the run Sphere.2D.$a_d$.4096.42.5.
\label{Sphere2DConserv}}
\end{figure}

Solving the system (\ref{EulerTr1}) - (\ref{EulerTr3}) via an operator
splitting technique inevitably leads to conservativity errors when the
solution is transformed back to the inertial frame. Therefore, it is
extremely important to assess the price in conservativity that is
being paid by using this method as well as possible ways to control
and minimize such errors.

It was discussed in \S~\ref{Accuracy} that the main cause of
conservativity violation are the errors $\delta \tilde q^{n+1}_S$ and
$\delta^2 \tilde q^{n+1}_{HS}$ (see eq. (\ref{SolutionError})).
Consider Fig.~\ref{Sphere2DConserv} which shows temporal evolution of
the relative error in total mass and energy conservation in the
highest resolution simulations of expansion of a non-rotating sphere
into vacuum. The overall behavior is rather similar in both runs. One
point should be noted, though, which is not as obvious due to the
logarithmic scale. The growth rate of the conservativity errors is
increasing in time, except for the error in total energy in the
delayed stretch run. That is the manifestation of the fact that the
errors $\delta \tilde q^{n+1}_S$ and $\delta^2 \tilde q^{n+1}_{HS}$
indicated in eq. (\ref{SolutionError}) accumulate at each time step
and the source term operator, acting on them at the subsequent step,
produces an even larger error. Therefore, it is very important to
minimize them. Aside from the brute force resolution increase, closely
adjusting the moving frame velocity to the fluid velocity is the only
method that does not penalize the performance and, instead, can cause
a significant speed up due to larger time steps in the computational
domain. Indeed, in the case of the delayed stretch run, which did not
have during the acceleration phase a significant velocity mismatch
characteristic of the run Sphere.2D.a.2048.50, the conservativity
error is two-three orders of magnitude lower. In fact, the error in
total energy conservation in case of the delayed stretch even shows
the decreasing trend. In both runs, as mentioned above, the magnitude
of the errors could have been reduced even further by decreasing the
target relative error in the implicit source term solver from the
value of $10^{-4}$ used in these simulations. By reducing that value
down to $10^{-12} - 10^{-13}$ we were able to decrease the
conservativity errors essentially to the order of machine
precision. On the other hand, that also resulted in about an order of
magnitude slower performance of the solver due to the much higher
number of iterations needed to provide the requested level of
accuracy. We also find that the use of Strang splitting can have a
much more profound effect on the conservativity of the solution than
on its accuracy. In particular, in this case the error in mass and
total energy conservation can be additionally decreased by about 2
orders of magnitude for each resolution.

\begin{figure}[t]
\epsscale{1.08}
\plottwo{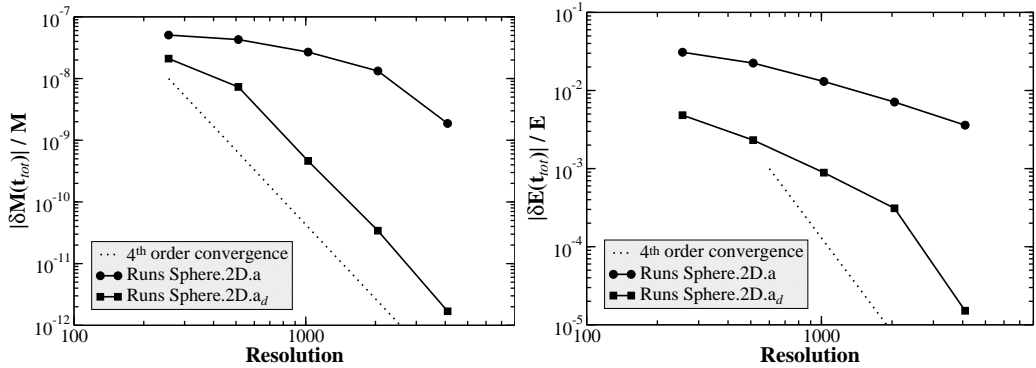}{fig9b.eps}
\caption{Dependence of the final relative error in total mass (\emph{left})
and total energy (\emph{right}) conservation on resolution for the
simulations of expansion of a non-rotating sphere into vacuum.
\label{Sphere2DConservConverg}}
\end{figure}

Figure~\ref{Sphere2DConservConverg} shows that the use of the delayed
stretch consistently results in much lower conservation errors at all
resolutions. Moreover, the use of the delayed stretch not only
significantly reduces the magnitude of the error, but also increases
the rate at which it drops with increasing resolution. The dashed line
in both figures indicates the rate of the conservation error decrease
proportional to the inverse $4^{th}$ power of the resolution. That
would correspond to the ideal rate of decrease in the case when the
only contribution to the conservation error is due to the term $\delta
\tilde q^{n+1}_S$ in eq. (\ref{SolutionError}), which drops as the
$4^{th}$ power of the spatial and temporal step. As previously
discussed, such situation would correspond to the case when the error
$\delta \tilde q^{n+1}_{HS}$ due the hydrodynamic solver becomes very
small, thus essentially removing the seed for the second contributor
to the conservation error, namely the error $\delta^2 \tilde
q^{n+1}_{HS}$. Indeed, such behavior is characteristic only of the
delayed stretch runs.

Finally, it should be noted that, as was discussed in
\S~\ref{EqnInvariance}, given the choice (\ref{BestChoice}) of
scaling parameters $\alpha$ and $\beta$ it is possible to always
conserve mass. Moreover, in the tests considered above momentum would
also be conserved since the runs were performed in a constant velocity
frame.

\subsection{Test results: method convergence}
%-------------------------------------------------------------------------------
\label{Convergence}

\begin{figure}
\epsscale{1.08}
\plottwo{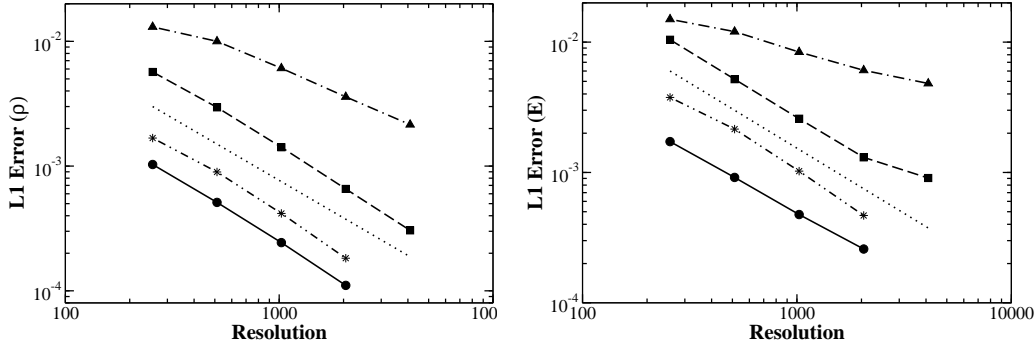}{fig10b.eps}
\caption{Convergence of the $L1$ error in density (\emph{left}) and
total energy (\emph{right}) for the simulations of expansion of a
non-rotating sphere into vacuum. Dotted lines show the expected rate
of convergence. \emph{Left:} \emph{Circles}: convergence of runs
Sphere.2D.S.256 - 2048 to the run Sphere.2D.S.4096 at $ t = 0.024$
(order of convergence $O(1.04)$); \emph{Squares}: convergence of runs
Sphere.2D.a.128 - 2048.50 to the run Sphere.2D.S.4096 at $ t = 0.024$
($O(1.0)$); \emph{Triangles}: convergence of runs Sphere.2D.a.128 -
2048.50 to the run Sphere.2D.$\textrm{a}_{d}$.4096.42.5 at $ t = 0.75$
($O(0.6)$); \emph{Stars}: convergence of runs
Sphere.2D.$\textrm{a}_{d}$.256 - 2048.42.5 to the run
Sphere.2D.$\textrm{a}_{d}$.4096.42.5 at $t = 0.75$
($O(0.98)$). \emph{Right:} Order of convergence: \emph{Circles} -
$O(0.92)$; \emph{Squares} - $O(0.99)$; \emph{Triangles} - $O(0.42)$;
\emph{Stars} - $O(0.92)$.
\label{Sphere2DConvergence}}
\end{figure}

Another crucial indicator of the quality of a numerical method is its
convergence properties. Fig.~\ref{SedovConvergence} shows the
convergence of the numerical solution to the exact one in simulations
of the Sedov blast wave. It can be seen that the convergence rate is
first-order as would be expected for the case of a discontinuous flow.
Moreover, the convergence rate is fairly insensitive to the type of
the moving frame used. Interestingly, the total energy shows the
convergence rate that is higher than that of density.

\begin{figure}[t]
\epsscale{1.02}
\plottwo{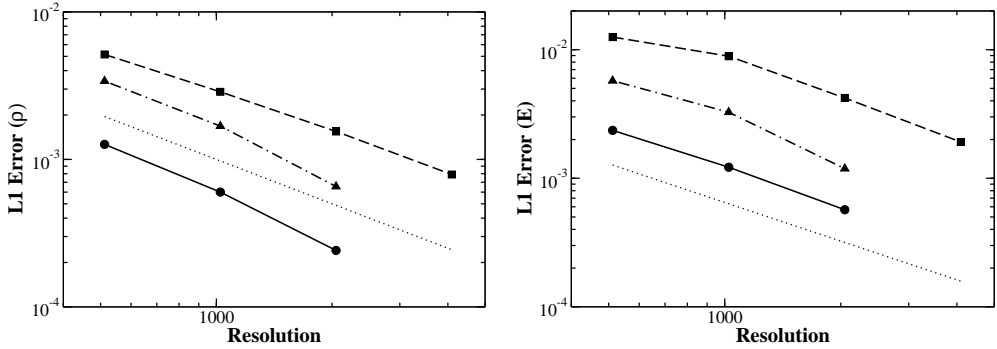}{fig11b.eps}
\caption{Convergence of the $L1$ error in density (\emph{left}) and
total energy (\emph{right}) for the simulations of expansion of a
rotating sphere into vacuum. Dotted lines show the expected rate of
convergence. \emph{Left:} \emph{Circles}: convergence of runs
SphereRot.2D.S.512 - 2048.80 to the run SphereRot.2D.S.4096 at $ t =
0.02$ (order of convergence $O(1.13)$); \emph{Squares}: convergence of
runs SphereRot.2D.$\textrm{e}_2$.256 - 2048.80 to the run
SphereRot.2D.S.4096 at $ t = 0.02$ ($O(0.87)$); \emph{Triangles}:
convergence of runs SphereRot.2D.$\textrm{e}_2$.256 - 1024.80 to the
run SphereRot.2D.$\textrm{e}_2$.2048.80 at $t = 0.75$ ($O(1.1)$).
\emph{Right:} Order of convergence: \emph{Circles} - $O(0.99)$;
\emph{Squares} - $O(0.76)$; \emph{Triangles} - $O(0.98)$.
\label{Sphere2DRotConvergence}}
\end{figure}

We also consider the convergence of the solutions obtained in the
simulations of expansion of a non-rotating and a rotating sphere into
vacuum at the early stage of system evolution, i.e., at the end of the
acceleration phase, as well as at the later stage when the flow
expansion is quite substantial. The results are shown in Figs.
\ref{Sphere2DConvergence} - \ref{Sphere2DRotConvergence}, giving 
the convergence of the $L1$ error in density and total energy. Note,
that the resulting $L1$ error was normalized to the maximum value of
the given quantity $q_{p,max}$ in the solution profile of the run
being considered in order to allow error comparison at early and late
stages of system evolution. Since the reference run always has the
highest resolution, its diagonal cut was interpolated individually to
the points of each run, being studied for convergence.

The reference ``exact'' solution at the end of the acceleration phase
was taken based on the highest resolution run performed in the
inertial reference frame, i.e., the run Sphere.2D.S.4096 in the
non-rotating case and the run SphereRot.2D.S.4096 in the rotating
case\footnote{Note that points that correspond to the moving frame
runs, except for the points marked with stars showing the delayed
stretch runs, indicate the resolutions that are twice higher than the
ones at which the runs were performed. That is done since the initial
cell size of the moving frame runs was the same as that of the
corresponding inertial frame runs, as was discussed in
\S~\ref{TestTypes}, however, their domain extent was twice smaller,
hence the twice smaller number of cells per dimension in the
domain. Therefore, in the figures they were indicated as having twice
the number of cells in order to maintain the correspondence with the
inertial frame runs.}. The reference ``exact'' solution in the case of
a non-rotating sphere at $t_{tot} = 0.75$, i.e., when the sphere has
expanded by almost two orders of magnitude, was the highest resolution
run performed with the delayed stretch, namely the run
Sphere.2D.$\textrm{a}_d$.4096.42.5. The reference ``exact'' solution
in the case of a rotating sphere at $t_{tot} = 0.75$ was the highest
resolution run performed in the expanding and rotating reference
frame, i.e., the run SphereRot.2D.$\textrm{e}_2$.4096.80. The caption
of each figure indicates the order of convergence for each curve. The
convergence in all cases is first-order. The only exception is the
lower convergence rate at a later stage of the runs performed without
the delayed stretch, i.e., the dash-dotted line. The reason for that
is the contamination of the solution by a large error due to the
significant mismatch between the velocities of the global flow and the
reference frame during the acceleration phase, as was discussed in
\S~\ref{Accuracy} and \S~\ref{Conservativity}.
Fig.~\ref{Sphere2DRotConvergence} shows that similar results are
produced in the case of a rotating sphere as well. Ideally, the flow
in such system should be smooth. Consequently the use of the
second-order scheme should lead to the second-order convergence. In
practice, however, there is always a discontinuity present in the
problem as the result of the propagation of the sphere material into
the ambient medium. In an ideal situation ambient density would have
to be set at zero which is impossible in a Eulerian code. Therefore,
even though the ambient density can be very small it still produces a
discontinuity thereby decreasing the order of convergence of the
overall solution.

\begin{figure}[t]
\epsscale{1.02}
\plottwo{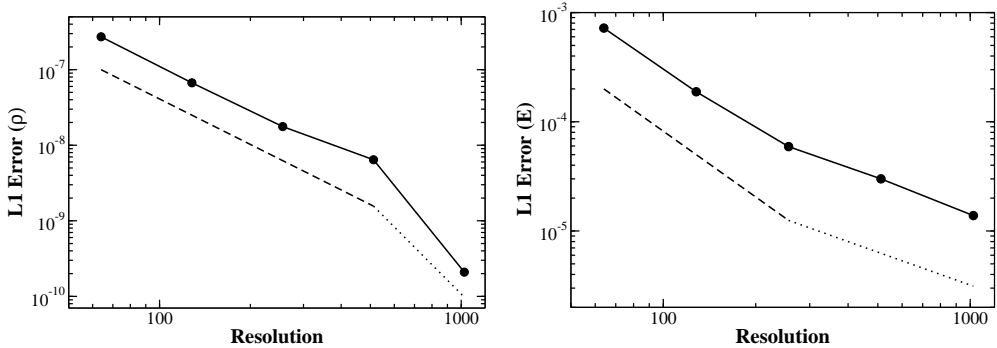}{fig12b.eps}
\caption{Convergence of the $L1$ error in density (\emph{left})
and total energy (\emph{right}) in runs Clump.2D.a.*.100. Dashed line
in both panels corresponds to second-order convergence, while the
dotted line in the left panel corresponds to fourth-order convergence
and in the right panel - to first-order convergence. Overall
convergence order in density is $O(2.01)$ and total energy is
$O(1.88)$.
\label{ClumpConvergence}}
\end{figure}

Fig.~\ref{ClumpConvergence} shows the convergence of the $L1$ error in
density and total energy in the simulations of isentropic expansion of
a uniform pressure field with the embedded density structure. In this
case second-order convergence of the numerical solution to the exact
one is achieved. Moreover, the density discontinuity remains sharp
throughout the duration of the simulations. The convergence rate in
total energy for the two highest resolution simulations drops to first
order due to the fact that the errors in density and pressure become
small enough so that the overall solution in total energy becomes
affected by the error arising as a result of the transformation of the
solution to the laboratory reference frame.

\section{Discussion and Conclusions}
%-------------------------------------------------------------------------------
\label{Discussion}

We have presented the method for computation of fluid flows
characterized by the following two properties: (1) large degree of
expansion/contraction on the evolutionary timescale; (2) domination of
the flow velocity field by the global component associated with
expansion/contraction and/or rotation of the flow. The key of the
method is the transformation $\mathbf{\Lambda}$
(eq. (\ref{Transform})) of the computational domain to a non-inertial
reference frame that is comoving with the global flow. Consequently,
the fluid variable fields, including velocity, are transformed
according to the eqs. (\ref{newvel}), (\ref{rhotr}), (\ref{Ptr}). The
formulation based on such general scaling transformation of the fluid
variable fields provides a large degree of flexibility in terms of the
form of the source terms which can be suited to the needs of a
particular problem. Transition to a moving frame allows to accommodate
naturally large changes of the flow extent. Moreover, treatment of
rapidly rotating flows, e.g., compact stars and stellar cores, is
naturally incorporated in this framework. We also showed that the
conservative formulation of the equations of fluid dynamics exists
only in the case of restricted reference frame transformations, namely
only for non-rotating reference frames expanding/contracting with a
constant velocity, and only for the polytropic index $\gamma = 1 +
\sfrac{2}{\nu}$.

\begin{figure}[t]
\epsscale{0.55}
\plotone{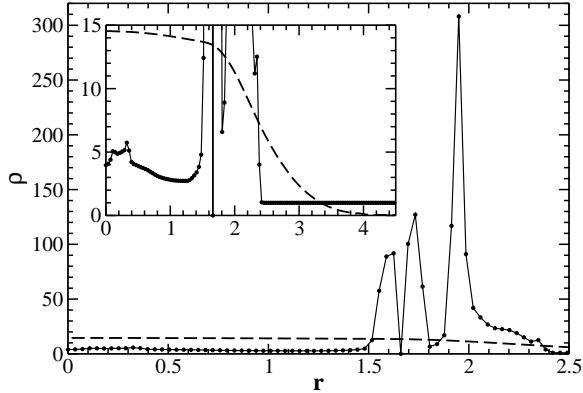}
\caption{Test Sphere.2D.a.256.50 performed with Zeus-MP (solid line).
Shown is the diagonal cut through the domain at $t = 0.12$. Dashed
line shows the reference solution obtained with the method presented
here. Inset shows the zoom-in of the main graph.
\label{Zeus_vs_Alla}}
\end{figure}

The first key advantage of this approach over other methods that can
be used for computation of such flows, namely the AMR and traditional
moving mesh techniques, is the fact that thermal and local kinetic
energies are comparable in magnitude in the moving reference frame,
thereby eliminating the high Mach number problem. All moving mesh
methods operate in the inertial frame $\mathbf{X}$, moreover
practically all of them are designed to evolve the total energy of the
flow. To illustrate the consequences such approaches have on the
solution we performed the test Sphere.2D.a.256.50 involving expansion
of a non-rotating sphere into vacuum with the Zeus-MP code (v. 2.0.2)
\citep{Zeus}. The problem setup was identical to the one described in
\S~\ref{TestTypes} with one exception. In our simulations ambient
material is set to be co-expanding with the reference frame in order
to minimize its dynamical effects on the expanding material of the
sphere. Setting ambient material to co-expand with the computational
grid proved to be impossible in our Zeus-MP simulation since the
overwhelming dominance of the total energy of the ambient material by
its kinetic component causes an immediate breakdown of the
solution. Therefore, we set ambient material to be stationary in the
laboratory frame. The grid motion was prescribed to be identical to
the grid motion in our simulations and outflow boundary conditions
were used on outer domain boundaries, while perfectly reflective
boundary conditions were used on the boundaries containing the fixed
point of expansion. The run was performed with CFL = 0.5 and
artificial viscosity parameter was set to 2.0. Result of the
simulation is shown in Fig.~\ref{Zeus_vs_Alla}. The solid line shows
the density distribution along the diagonal cut of the domain in our
Zeus-MP run, while the dashed line shows for reference the result
obtained in the simulation using the method presented here. The
complete breakdown of the solution can be seen in the figure. As the
outer layers start to expand and their velocity increases thereby
rapidly increasing the fraction of the kinetic part of total energy
the pressure structure of the rarefaction wave starts to degrade
dramatically. The accumulating errors in pressure propagate inward
with the rarefaction wave very rapidly completely destroying the
solution. All of moving mesh techniques that evolve total energy would
suffer from the same problem, albeit to a different extent depending
on spatial and temporal accuracy of their hydrodynamic scheme. There
exist moving mesh algorithms that evolve internal energy. In
particular, Zeus-MP is capable of operating in such a mode and in that
case it produces results comparable to the ones obtained with the
method presented in this work. Such schemes are also non-conservative,
however, unlike the method presented here there are no means in them
to control the magnitude of the conservativity error besides the brute
force method of resolution increase. Moreover the conservativity error
would converge with at most the order of the scheme, e.g., second
order, while in our method it is possible to achieve fourth-order
convergence of the conservativity error, as was shown in
\S~\ref{Conservativity}. 

It is also important to emphasize that moving mesh algorithms,
especially the ones that evolve internal energy, are tightly coupled
to a particular implementation of a hydrodynamic scheme which may not
be appropriate for all problems when, for example, a higher order
scheme might be required. The key practical advantage of the method
presented here is the form-invariance of the homogeneous part of fluid
equations. Consequently, eqs. (\ref{EulerTr1}) - (\ref{EulerTr3})
allow for their very easy implementation in a numerical code since all
of the effects associated with the moving frame are provided for by
source terms. The latter can be solved via operator splitting
techniques and there exist very accurate and efficient methods for
solving such systems of ordinary differential equations. Unlike the
case of the moving mesh formulation (\ref{MovMeshLaw}) which redefines
the state vector and the flux functions and thus could require major
changes in the hydrodynamic solvers of numerical codes, the
formulation discussed here does not affect the hydrodynamic solvers.
Thus it allows essentially for the implementation of one general
source term integrator that can then be easily combined with any
Eulerian hydrodynamic scheme and AMR strategy and, therefore, can
utilize the wide potential of existing techniques for solving systems
of hyperbolic equations. In particular, we have demonstrated
performance of the method in combination with the second-order
dimensionally split hydrodynamic scheme used with cell-based AMR (Alla
code) and the second-order unsplit scheme using grid-based AMR
(AstroBEAR code). Such portability of the method may be especially
beneficial in combining this approach with more involved hydrodynamic
schemes, in particular implicit and low Mach number solvers for which
there may exist no immediate implementation for the solution of the
equations similar to eq. (\ref{MovMeshLaw}) and which may provide
further improvements in efficiency in problems in which the local
velocity component is small, i.e., $\tilde u_i < \tilde c_i$.

We conducted extensive numerical testing of the method in order to
address three key issues: (1) what is the solution accuracy in the
``worst-case scenario'' applications, i.e., problems for which this
method is poorly suited; (2) how different types of reference frame
motions correlate with the solution accuracy; (3) what is the
performance of the method in the case of its optimal applications. The
goal of the first two questions was to determine the limitations in
applicability of this method as well as the requirements that have to
be imposed on the feedback mechanism filters, used in realistic
applications to generate reference frame motions, which will not lead
to the degradation of solution accuracy.

In tests based on the strong point explosion, which represented an
application not optimal for this method, solution accuracy and rate of
convergence to the exact solution were generally better than, or
similar to, that of the reference solution obtained in the stationary
reference frame. Moreover, using the appropriate timestep limiting,
given by eq. (\ref{dtLimited}), it was possible to have both low
frequency high amplitude and high frequency low amplitude reference
frame oscillations without any loss of solution accuracy. The test
problem involving converging shock demonstrated accuracy of the method
also in the case of contracting reference frames.  Typically, given a
source term solver which has higher accuracy than the hydrodynamic
solver, the latter completely determines the overall solution accuracy
and convergence properties. In our tests of isentropic expansion of a
uniform pressure field with an embedded density structure we observe
convergence rate that would be expected for the 2nd-order accurate
code. On the other hand, the error in conservativity is completely
determined by the source term accuracy. Indeed, in tests involving
expansion of a non-rotating and a rotating sphere into vacuum we
observed the 4th-order convergence of the total error in conservation
of mass and energy. In general, the key factor that determines the
accuracy and conservativity of the solution is how well the frame and
fluid velocities are correlated.

In summary, the method presented here provides excellent performance
in computation of fluid flows described above, in particular due to
much lower errors in pressure. The method is able to accommodate a
large range of fluid motions, including the highly oscillatory ones,
without the loss of accuracy. Finally, the non-conservative
formulation of the equations, though, as was shown, unavoidable due to
the non-inertiality of the computational frame, virtually does not
affect the accuracy or convergence properties of the solution.

{\bf Acknowledgments} This work was supported in part by the US
Department of Energy under contract B523820 to the Center for
Astrophysical Thermonuclear Flashes at the University of Chicago.  AYP
also expresses gratitude to the Institute for Pure and Applied
Mathematics (IPAM), UCLA for support during the semester program
``Grand Challenge Problems in Computational Astrophysics'', March -
June, 2005. The authors are grateful to anonymous referees for
valuable comments and criticisms.

%-------------------------------------------------------------------------------
\appendix
\section{Case of Cylindrical Symmetry}
%-------------------------------------------------------------------------------
\label{Cylindrical}

Here we give the set of transformed Euler equations in the
expanding / contracting reference frame in the case of cylindrical
symmetry. We consider the transformed equations in the absence of
rotation around the symmetry axis of the coordinate system, i.e.,
$u_\phi = 0$. As before, we consider the transformation
$\mathbf{\Lambda_{\{r,z\}}}$ of the inertial frame
$\mathbf{X_{\{r,z\}}}=\{r,z,t\}$, defined in cylindrical coordinates,
to a comoving expanding/contracting reference frame $\mathbf{\tilde
X_{\{\tilde r, \tilde z\}}}=\{\tilde r, \tilde z, \tau \}$
\beq
\mathbf{\Lambda_{\{r,z\}}} = \left\{ \begin{array}{lcl}
\tilde r  & = &  a^{-1}(t)r, \\
\tilde z  & = &  a^{-1}(t)z, \\
\tau      & = & \displaystyle \int_0^t \frac{dt}{a^{\beta+1}(t)}.
\end{array} \right.
\label{TransformCyl}
\eeq
Transformation $\mathbf{\Lambda_{\{r,z\}}}$ and the expansion
coefficient $a(t)$ have the same properties as before. Decomposition
of the velocity field $\vect{u}$ is similar to eq. (\ref{newvel})
\beq
\left\{ \begin{array}{rcl}
u_r & \, = \, & \displaystyle a^{-\beta}\left(\frac{d\ln a}{d\tau} \tilde r
+ \tilde u_r\right), \\
u_z & \, = \, & \displaystyle a^{-\beta}\left(\frac{d\ln a}{d\tau} \tilde z
+ \tilde u_z\right).
\end{array} \right.
\label{newvelcyl}
\eeq
Density, pressure, and internal energy fields are transformed as
before (eqs. (\ref{rhotr}), (\ref{Ptr})).

Euler equations in cylindrical coordinates in the reference frame
$\mathbf{X_{\{r,z\}}}$ are
\begin{eqnarray}
\pd{\rho}{t} + \pd{\left( \rho u_r \right)}{r}
+ \pd{\left( \rho u_z \right)}{z} & \, = \, & -\frac{\rho u_r}{r},
\label{EulerCyl_rho} \\
\pd{\left( \rho u_r \right)}{t} + \pd{\left( \rho u^2_r \right)}{r}
+ \pd{\left( \rho u_r u_z \right)}{z} + \pd{P}{r}
& \, = \, & -\frac{\rho u^2_r}{r},
\label{EulerCyl_vr} \\
\pd{\left( \rho u_z \right)}{t} + \pd{\left( \rho u_r u_z \right)}{r}
+ \pd{\left( \rho u^2_z \right)}{z}  + \pd{P}{z}
& \, = \, & -\frac{\rho u_r u_z}{r},
\label{EulerCyl_vz} \\
\pd{E}{t} + \pd{u_r\left( E + P \right)}{r}
+ \pd{u_z\left( E + P \right)}{z} & \, = \, & -\frac{u_r\left( E + P \right)}{r}.
\label{EulerCyl_E}
\end{eqnarray}

The transformed Euler equations in the frame $\mathbf{\tilde
X_{\{\tilde r, \tilde z\}}}$ have the form
\begin{eqnarray}
\pd{\tilde\rho}{\tau}
+ \pd{\left( \tilde\rho \tilde u_r \right)}{\tilde r} + \pd{\left(
\tilde\rho \tilde u_z \right)}{\tilde z} & \, = \, & \left(\alpha - 3 \right)
\frac{d\ln a}{d\tau} \tilde\rho - \frac{\tilde\rho \tilde u_r}{\tilde r}.
\label{Ctcyl} \\
\pd{\left( \tilde\rho \tilde u_{\tilde r} \right)}{\tau}
+ \pd{\left( \tilde\rho \tilde u^2_{\tilde r} \right)}{\tilde r}
+ \pd{\left( \tilde\rho \tilde u_{\tilde r} \tilde u_{\tilde z} \right)}{\tilde z}
+ \pd{\tilde P}{\tilde r}
& \, = \, & \left(\alpha + \beta - 4 \right)\frac{d\ln a}{d\tau}
  \tilde \rho \, \tilde u_{\tilde r} - \nonumber \\
& & \Bigg\{\frac{d^2\ln a}{d\tau ^2} - \beta\left( \frac{d\ln a}{d\tau}\right)^2\Bigg\}
  \tilde\rho \, \tilde r - \nonumber \\
& & \frac{\tilde\rho \tilde u^2_{\tilde r}}{\tilde r},
\label{Mrtcyl} \\
\pd{\left( \tilde\rho \tilde u_{\tilde z} \right)}{\tau}
+ \pd{\left( \tilde\rho \tilde u_{\tilde r} \tilde u_{\tilde z} \right)}{\tilde r}
+ \pd{\left( \tilde\rho \tilde u^2_{\tilde z} \right)}{\tilde z}
+ \pd{\tilde P}{\tilde z}
& \, = \, & \left(\alpha + \beta - 4 \right)\frac{d\ln a}{d\tau}
  \tilde \rho \, \tilde u_{\tilde z} - \nonumber \\
& & \Bigg\{\frac{d^2\ln a}{d\tau ^2} - \beta\left( \frac{d\ln a}{d\tau}\right)^2\Bigg\}
  \tilde\rho \, \tilde z - \nonumber \\
& & \frac{\tilde\rho \tilde u_{\tilde r} \tilde u_{\tilde z}}{\tilde r},
\label{Mztcyl} \\
\pd{\tilde E}{\tau} + \pd{\tilde u_{\tilde r}\left( \tilde E + \tilde P \right)}{\tilde r}
+ \pd{\tilde u_{\tilde z}\left( \tilde E + \tilde P \right)}{\tilde z}
& \, = \, & \frac{d\ln a}{d\tau} \bigg[\left(\alpha + 2\beta - 3 \right) \tilde E
- 3 \tilde P -\tilde \rho \, \tilde V^2 \bigg] - \nonumber \\
& & \Bigg\{\frac{d^2\ln a}{d\tau ^2} - \beta\left( \frac{d\ln a}{d\tau}\right)^2\Bigg\}
  \tilde\rho \, \left( \tilde u_{\tilde r} \, \tilde r + \tilde u_{\tilde z} \,
  \tilde z \right) - \nonumber \\
& & \frac{\tilde u_{\tilde r}\left( \tilde E + \tilde P \right)}{\tilde r},
\label{Etcyl}
\end{eqnarray}
where $\tilde V^2 = \tilde u^2_{\tilde r} + \tilde u^2_{\tilde z}$. As
it can be seen, the contribution of cylindrical symmetry to the
original set of the transformed Euler equations (\ref{EulerTr1}) -
(\ref{EulerTr3}) in the computational frame in the absence of rotation
is manifested by the addition of the geometric source terms, which are
of the same form as in the inertial frame but which include the local
velocity field in the computational frame, and by the value of the
dimensionality parameter $\nu = 3$. The latter is explained by the
fact that, even though the problem is described in the two-dimensional
space, its effective dimensionality in the case of cylindrical
symmetry is 3.

%-------------------------------------------------------------------------------

\end{document}